\documentclass[useAMS,usenatbib]{mn2e}
\def\stis{{\rm{STIS}}~}
 
\def\hst{{\it{HST}}} 
\def\sp{{\it{Spitzer}}} 
\def\ir{{\rm{IRAC}}~} 

\newcommand{\bjdutc}{\ensuremath{\rm {BJD_{UTC}}}}
\newcommand{\bjdtt}{\ensuremath{\rm {BJD_{TT}}}}
\newcommand{\bmjdobs} {\rm {BMJD\_OBS}}

\def\wp{\mbox{WASP-6}}
\def\tp{\mbox{\it{(T-P)}}~}
\def\hdtwo{\mbox{HD\,209458b}} 
\def\hdone{\mbox{HD\,189733b}}

\def\sp{\mbox{\it{Spitzer}}} 
\def\ir{\mbox{IRAC}}
\def\rps{\mbox{${\rm{R_p/R_\ast}}$}}  
\def\ar{$a/R_{\ast}$} 
\usepackage[pdftex]{graphicx}
\usepackage{amsmath}
\usepackage{amssymb}
\usepackage{deluxetable}
\usepackage[breaklinks,colorlinks,urlcolor=blue,citecolor=blue,linkcolor=blue]{hyperref}
\usepackage{epstopdf}
\usepackage{natbib}
\usepackage{subfigure}
\usepackage{verbatim}
\usepackage{tablefootnote}
\usepackage{caption}
\usepackage{graphicx}
\usepackage{hyperref}
\usepackage{bookmark}

\title[{\it{HST}}~STIS~transit spectroscopy of WASP-6b]
{ \textit{\textbf{HST}} hot-Jupiter transmission spectral survey: \\
Haze in the atmosphere of WASP-6b}

\author[N. Nikolov et al.]
 {N.~Nikolov, $^{1}$\thanks{ E-mail: nikolay@astro.ex.ac.uk (NN)}
  D.~K.~Sing,$^1$
  A.~S.~Burrows,$^2$
  J.~J.~Fortney,$^3$  
  G.~W.~Henry,$^4$
  F.~Pont,$^1$
  \newauthor 
  G.~E.~Ballester,$^5$
  S.~Aigrain,$^6$
  P.~A.~Wilson,$^{1,7}$
  C.~M.~Huitson,$^8$
  N.~P.~Gibson,$^9$   
  \newauthor  
  J.-M. D\'esert,$^8$
  A.~Lecavelier des Etangs,$^{7}$ 
  A.~P.~Showman,$^5$
  \newauthor  
  A.~Vidal-Madjar,$^{7}$
  H.~R.~Wakeford,$^1$ 
  K.~Zahnle$^{10}$\\
\\
  $^{1}$Astrophysics Group, School of Physics, University of Exeter, Stocker Road, Exeter EX4 4QL, UK\\
  $^{2}$Department of Astrophysical Sciences, Peyton Hall, Princeton University, Princeton, NJ 08544, USA\\ 
  $^{3}$Department of Astronomy and Astrophysics, University of California, Santa Cruz, CA 95064, USA\\ 
  $^{4}$Tennessee State University, 3500 John A. Merritt Blvd., PO Box 9501, Nashville, TN 37209, USA\\ 
  $^{5}$Lunar and Planetary Laboratory, University of Arizona, Tucson, AZ 85721, USA\\ 
  $^{6}$Department of Physics, University of Oxford, Denys Wilkinson Building, Keble Road, Oxford OX1 3RH, UK\\ 
  $^{7}$CNRS, Institut dÕAstrophysique de Paris, UMR 7095, 98bis boulevard Arago, 75014 Paris, France\\
  $^{8}$CASA, Department of Astrophysical and Planetary Sciences, University of Colorado, 389-UCB, Boulder, CO 80309, USA\\
  $^{9}$European Southern Observatory, Karl-Schwarzschild-Str. 2, D-85748 Garching bei M{\"u}nchen, 
 Germany\\ 
  $^{10}$NASA Ames Research Center, Moffett Field, CA 94035, USA\\
}

\begin{document}

\date{Accepted 2014 November 17.  Received 2014 November 17; in original form 2014 July 25 }

\pagerange{\pageref{firstpage}--\pageref{lastpage}} \pubyear{2013}

\maketitle

\label{firstpage}

\begin{abstract}

We report {\it{Hubble Space Telescope (HST)}} optical to near-infrared transmission spectroscopy of the hot Jupiter WASP-6b, measured with the Space Telescope Imaging Spectrograph (STIS) and  {\it{Spitzer's}} InfraRed Array Camera (\ir). The resulting spectrum covers the range $0.29-4.5\,\mu$m. We find evidence for modest stellar activity of \wp~and take it into account in the transmission spectrum. The overall main characteristic of the spectrum is an increasing radius as a function of decreasing wavelength corresponding to a change of $\Delta (\rps)=0.0071$ from 0.33 to $4.5\,\mu$m. The spectrum suggests an effective extinction cross-section with a power law of index consistent with Rayleigh scattering, with temperatures of $973\pm144$~K at the planetary terminator. We compare the transmission spectrum with hot-Jupiter atmospheric models including condensate-free and aerosol-dominated models incorporating Mie theory. While none of the clear-atmosphere models is found to be in good agreement with the data, we find that the complete spectrum can be described by models that include significant opacity from aerosols including  Fe-poor Mg$_2$SiO$_4$, MgSiO$_3$, KCl and Na$_2$S dust condensates. WASP-6b is the second planet after HD\,189733b which has equilibrium temperatures near $\sim1200$\,K and shows prominent atmospheric scattering in the optical.

\end{abstract}


\section{Introduction}\label{sec:introsec}
Multi-wavelength transit observations of hot-Jupiters provide a unique window to the chemistry and structure of the atmospheres of these distant alien worlds. During planetary transits, a small fraction of the stellar light is transmitted through the planetary atmosphere and signatures of atmospheric constituents are imprinted on the stellar spectrum \citep{seager00}. Transmission spectroscopy, where the planet radius is measured as a function of wavelength, has revolutionised our understanding of extrasolar gas-giant atmospheres. Numerous studies from both space and ground-based observations have led to the characterisation and detection of atomic and molecular features as well as hazes and clouds in the atmospheres of several hot Jupiters revealing a huge diversity \citep{charbonneau02, 
redfield08, 
grillmair08, 
snellen08, snellen10a,
pont08,  pont13, 
sing11b,sing12,sing13, 
brogi12, 
bean13,  
deming13,
huitson13, 
stevenson14,
nikolov14, 
wakeford13, 
gibson13a, gibson13b, 
birkby13}.

Atmospheric hazes have currently been detected in hot-Jupiter exoplanets over a wide range of atmospheric temperature regimes including HD 189733b and WASP-12b representing the cool and hot extremes for hot Jupiters at equilibrium temperatures around 1200 and 3000 K, respectively. Hazes are considered to originate from dust condensation (forming aerosols in the atmospheres of exoplanets) or a result from photochemistry \citep{marley13}. Transmission spectroscopy studies may constrain the most probable condensates responsible for these two planets, with iron-free silicates (e.g. MgSiO$_3$) and corundum (Al$_2$O$_3$) being two prime candidates for the above temperature regimes \citep{lecavelier08a, sing13}. Cloud-free hot-Jupiter atmospheric models predict sodium and potassium to be the dominant absorbing features in optical transmission spectra \citep{seager00, fortney05a}. 

In this paper we present new results for \wp b from a large \hst~hot-Jupiter transmission spectral survey comprising eight transiting planets. The ultimate aim of the project is to explore the variety of hot-Jupiter atmospheres, i.e. clear/hazy/cloudy, delve into the presence/lack of TiO/alkali and other molecular features, probe the diversities between possible subclasses and perform comparative exoplanetology. Initial results from four exoplanets have been presented so far in \cite{huitson13} for WASP-19b, \cite{wakeford13} and \cite{nikolov13} for HAT-P-1b and \cite{sing13} for WASP-12b and Sing et al. (2014) for WASP-31b. Already a wide diversity among exoplanet atmospheres is observed. In this paper we report new \hst\,optical transit observations with the STIS instrument and combine them with \sp~\ir\,broad-band transit photometry to calculate a near-UV to near-infrared transmission spectrum, capable of detecting atmospheric constituents. We describe the observations in Section\,\ref{sec:observationsec}, present the light curve analysis in Section\,\ref{sec:analysis}, discuss the results in Section\,\ref{sec:discussionsec} and conclude in Section\,\ref{sec:concl}.


\subsection{The \wp b system}\label{sec:introwasp}
Discovered by the Wide Angle Search for Planets, \wp\,b is an inflated sub-Jupiter-mass transiting extrasolar planet orbiting a moderately bright ${\rm{V}}= 11.9$ solar-type, mildly metal-poor star (with $T_{{\rm{eff}}}=5375\pm65$\,K, $\log{g}=4.61\pm0.07$ and [Fe/H]$=-0.20\pm0.09$) located in the southern part of the constellation  Aquarius \citep{gillon09a, doyle13}. The planet is moving on an orbit with a period of ${\rm{P\simeq3.6}}$\,days and semimajor axis $a\simeq0.042$ AU. The sky-projected angle between the stellar spin and the planetary orbital axis has been determined through observations of the Rossiter-McLaughlin effect by \cite{gillon09a} indicating a good alignment (${\rm{\beta=11^{+14}_{-18}~deg}}$) that consequently favours a planet migration scenario via the spin-orbit preserving tidal interactions with a protoplanetary disk. \cite{dragomir11} refined the \wp~system parameters and orbital ephemeris from a single ground-based transit observation finding good agreement between their results and the discovery paper. Although \cite{gillon09a} claimed evidence for non-zero orbital eccentricity, an analysis with new radial velocity data from \cite{husnoo12} brought evidence for non-significant eccentricity. Finally, \cite{doyle13} refined the spectroscopic parameters of \wp b's host star.

A cloud-free hot-Jupiter theoretical transmission spectrum predicts strong optical absorbers, dominated by Na\,{\sc i} and K\,{\sc i} absorption lines and H$_2$ Rayleigh scattering for a planetary system with the physical properties of \wp b, including an effective planetary temperature (assuming zero albedo and $f=1/4$ heat redistribution) of $~1194$\,K and surface gravity $g\simeq8~{\rm{m/s^2}}$ \citep{fortney08, burrows10}. The atmosphere of \wp b has recently been probed in the optical regime from the ground by \cite{Jordan13}. The authors found evidence for an atmospheric haze characterised by a decreasing  apparent planetary size with wavelength and no evidence for the pressure broadened alkali features. 

\section{Observations and Calibrations}\label{sec:observationsec}

\subsection{\hst~\stis\,spectroscopy}
\begin{figure}
        \centering
                 \includegraphics[trim = 14mm 0mm 5mm 0mm, clip, scale=0.37]{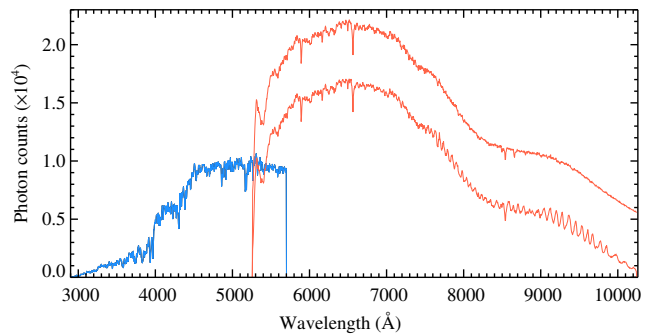}
         \caption{Typical \stis G430L and G750L spectra (blue and red continuos lines, respectively). A fringe-corrected spectrum (offset by $5 \times 10^3$ counts) is portrayed above the uncorrected red spectrum for comparison.}
\label{fig:stis_spec}
\end{figure}

We acquired low-resolution ($R= \Delta  \lambda / \lambda = 500-1040$) \hst~\stis spectra (Proposal ID GO-12473, P.I., D. Sing) during three transits of \wp\,b on UT~2012~June~10 (visit 9) and 16 (visit 10) with grating G430L ($\sim2.7$~\AA/pixel) and 2012 July~23 (visit 21) with grating G750L ($\sim4.9$~\AA/pixel). When combined, the blue and red \stis data provide complete wavelength coverage from 2900~{\AA} to 10\,300~{\AA} with a small overlap region between them from 5240 {\AA} to 5700~{\AA} (Fig.~\ref{fig:stis_spec}). Each visit consisted of five $\sim 96\ {\rm{min}}$ orbits, during which data collection was truncated with $\sim45\ {\rm{min}}$ gaps due to Earth occultations. Incorporating wide $52\times2''$ slit to minimise slit light loses and an integration time of $278~{\rm{s}}$ a total of 43 spectra were obtained during each visit. Data acquisition overheads were minimised by reading-out a reduced portion of the CCD with a size of $1024\times128$. This observing strategy has proven to provide high signal to noise ratio (S/N) spectra that are photometrically accurate near the Poisson limit during a transit event \citep{brown01, sing11b, huitson12, sing13}. The three \hst~visits were scheduled such that the second and third spacecraft orbits occurred between the second and third contacts of the planetary transit in order to provide good sampling of the planetary radius while three orbits secured the stellar flux level before and after the transit.

The data reduction and analysis process is somewhat uniform for the complete large \hst~program and follows the general methodology detailed in \cite{huitson13}, \cite{sing13} and \cite{nikolov14}. The raw \stis data was reduced (bias-, dark- and flat-corrected) using the latest version of the {\tt{CALSTIS}}\footnote{{\tt{CALSTIS}} comprises software tools developed for the calibration of STIS data \citep{katsanis98} inside the {\tt{IRAF}} (Image Reduction and Analysis Facility; http://iraf.noao.edu/) environment.} pipeline and the relevant up-to-date calibration frames. Correction of the significant fringing effect was performed for the G750L data using contemporaneous fringe flat frame obtained at the end of the observing sequence and the procedure detailed in \cite{goudfrooij98b} (Fig.~\ref{fig:stis_spec}).

Due to the relatively long \stis integration time (i.e. 287~s), the data contains large number of cosmic ray events, which were identified and removed following the procedures described in \cite{nikolov13}. It was found that the total number of pixels affected by cosmic ray events comprise $\sim4\,\%$ of the total number of pixels of each {\tt{2D}} spectrum. In addition, we also corrected all pixels identified by {\tt{CALSTIS}} as ``bad" with the same procedure, which together with the cosmic ray identified pixels resulted in a total of $\sim11\,\%$ interpolated pixels. 

Spectral extractions were performed in {\tt{IRAF}} employing the {\tt{APALL}} procedure using the calibrated {\tt{.flt}} science files after fringe and cosmic ray correction. We performed spectral extraction with aperture sizes in the range 6.0 to 18.0 pixels with a step of 0.2. The best aperture for each grating was selected based on the resulting lowest light curve residual scatter after fitting the white light curves (see Section\,\ref{sec:analysis} for details). We found that aperture sizes 12.0, 12.2 and 8.6 pixels satisfy this criterium for visits 9, 10 and 21 respectively. We then placed the extracted spectra to a common Doppler corrected rest frame through cross-correlation to prevent sub-pixel wavelength shifts in the dispersion direction. A wavelength solution was obtained from the {\tt{x1d}} files from {\tt{CALSTIS}}. The \stis spectra were then used to extract both white-light spectrophotometric time series and custom wavelength bands after integrating the appropriate flux from each bandpass.

The raw STIS light curves exhibit instrumental systematics similar to those described by \cite{gilliland99} and \cite{brown01}. In summary, the major source of the systematics is related with the orbital motion of the telescope. In particular the {\sl{HST}} focus is known to experience  quite noticeable variations on the spacecraft orbital  time scale, which are attributed to thermal contraction/expansion (often referred to as the ``breathing effect'') as the spacecraft warms up during its orbital day  and cools down during orbital night \citep{hasan93, hasan94, suchkov98}. We take into account the systematics associated with the telescope temperature variations in the transit light curve fits by fitting a baseline function depending on various parameters (Sec.~\ref{sec:hst_analysis}). 


\subsection{\sp~\ir~data}
Photometric data was collected for \wp~during two transits of its planet on UT 2013 January 14 and 21 with the \sp~space telescope \citep{Werner04} employing respectively the 4.5 and 3.6\,$\mu m$ channels of the Infrared Array Camera (IRAC, \citealt{Fazio04}) as part of program 90092 (P.I. J.-M. Desert). Both observations started shortly before ingress with effective integration times of 1.92\,s per image and were terminated $\sim2.2$\,h after egress resulting in 8\,320 images (see Fig.~\ref{fig:spitzer_data}), which were calibrated by the \sp~pipeline (version S19.1.0) and are available in the form of Basic Calibrated Data ({\tt{.bcd}}) files.

After organising the data we converted the images from flux in mega-Jansky per steradian ($MJy~sr^{-1}$) to photon counts (i.e. electrons) using the information provided in the FITS headers. In particular we multiplied the images by the gain and individual exposure time (FITS HEADER key words SAMPTIME and GAIN) and divided by the flux conversion factor (FLUXCONV). Timing of each image was computed using the UTC-based Barycentric Julian Date (\bjdutc) from the FITS header keyword \bmjdobs, transforming these time stamps into Barycentric Julian Date based on the BJD terrestrial time (TT) using the following conversion $\bjdtt \sim \bjdutc + 66.184$ s \citep{eastman10}. This conversion is preferable as leap seconds are occasionally added to the \bjdutc~standard \citep{knutson12, todorov13}.

We performed outlier filtering for hot (energetic) or lower pixels in the data by following each pixel through time. This task was performed in two steps, first flagging all pixels with intensity $8\text{-}\sigma$ or more away from the median value computed from the 5 preceding and 5 following images. The values of these flagged pixels were replaced with the local median value. In the second pass  we flagged and replaced outliers above the $4\text{-}\sigma$ level, following the same procedure. The total fraction of corrected pixels was $0.26\,\%$ for the $3.6\, \mu m$ and 0.06\,$\%$ for the $4.5\,\mu m$ channel. 

We estimated and subtracted the background flux from each image of the time series. To do this we performed an iterative $3\text{-}\sigma$ outlier clipping for each image to remove the pixels with values associated with the stellar (point-spread function) PSF, background stars or hot pixels, created a histogram from the remaining pixels and fitted a Gaussian to determine the sky background.

We measured the position of the star on the detector in each image incorporating the flux-weighted centroiding method\footnote{as implemented in the IDL ${\tt{box\_centroider.pro}}$, provided on the \sp~home-page: \url{http://irsa.ipac.caltech.edu}} using the background subtracted pixels from each image for a circular region with radius 3 pixels centered on the approximate position of the star. While we could perform PSF centroiding using alternative methods such as fitting a two-dimensional Gaussian function to the stellar image, previous experiences with warm \sp~photometry showed that the flux-weighted centroiding method is either equivalent or superior \citep{beerer11, knutson12, lewis13}. The variation of the $x$ and $y$ positions of the PSF on the detectors were measured to be 0.20 and 0.21 for the 3.6 $\mu m$ channel and 0.89 and 0.8 pixels for the 4.5 $\mu m$ channel.

We extract photometric measurements from our data following two methods. First, aperture photometry was performed with {\tt{IDL}} routine {\tt{APER}} using circular apertures ranging in radius from 1.5 to 3.5 pixels in increments of 0.1. We filtered the resulting light curves for $5\text{-}\sigma$ outliers with a width of 20 data points. 

We also performed photometry with time-variable aperture with a size scaled by the value of a quantity known as the noise pixel parameter \citep{mighell05, knutson12}, which is described in Section 2.2.2 of the IRAC instrument handbook and has been used in previous exoplanet studies to improve the results of warm \sp~photometry \citep[see e.g.][]{ charbonneau08, knutson12, todorov13, orourke14}. The noise parameter depends on the FWHM of the stellar point-spread function squared and is defined as:
\begin{equation}
\tilde{\beta} = \frac{(\sum_{i} I_i)^2}{ \sum_{i} I_i^2},
\end{equation}
where $I_i$ is the intensity detected by the $i$-th pixel. We use each image to measure the noise pixel parameter applying an aperture with radius of 4 pixels, ensuring that each pixel is considered should the border of the aperture cross that pixel. We extracted source fluxes from both channels with the aperture radii following the relation:

\begin{equation}
r = \sqrt{\tilde{\beta}} a_{0} + a_{1},
\end{equation}
where $a_{0}$ and $a_{1}$ were varied in the ranges 0.8 to 1.2 and -0.4 to +0.4 with step 0.1.  

The best results from both photometric methods were identified by examining both the residual root-mean-square (rms) after fitting the light curves from each channel as well as the white and red noise components measured with the wavelet technique detailed in \cite{carter09}. The second photometric method resulted in the lowest white and random red noise correlated with data points co-added in time or spectral sampling for the $3.6\,\mu m$ data with $a_{0} = 0.9$ and $a_{1} = -0.1$. For the $4.5\,\mu m$ data the first method gave better results with aperture radius 2.4 pixels and sky annulus defined between radii 2.40 and 6.48 pixels.

Finally, we also performed a visual inspection of the \sp~\ir~images for obvious stellar companions by aligning and stacking $\sim3000$ of the available images in both channels, finding no evidence for bright stellar sources within the $38\times 38$ arcsec$^2$ field of view.

\subsection{Stellar variability monitoring}\label{sec:stellaractivity}
Stellar activity can complicate the interpretation of exoplanet transmission spectra, especially when multi-instrument multi-epoch data sets are combined \citep{pont13}. As the star rotates active regions on its surface enter into and out of view, causing the measured flux to exhibit a quasi-periodic variability that introduces variation of the measured transit depth (as \rps). This becomes particularly important when transit observations made over several months or years are combined to construct an exoplanet transmission spectrum, as in our \hst/\sp~study.  

We obtained high-resolution spectra of \wp~from the publicly available HARPS data to search for evidence of chromospheric activity. All spectra in the wavelength region of the Ca\,{\sc ii}\,H\,\&\,K lines show evidence for emission, implying a moderate stellar activity compared to the same region of HARPS spectra obtained for HD189733. An examination of the \hst~and \sp~white and spectral light curves show no evidence for spot crossings. 

\begin{figure}
        \centering
                 \includegraphics[trim = 0mm 0mm 0mm 0mm, clip, scale=0.56]{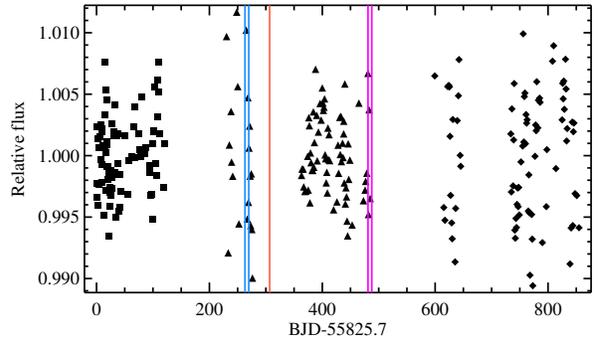}
         \caption{Fairborn Observatory Cousins $R$ filter light curve of \wp~from three seasons (indicated with different symbols). The vertical lines indicate the STIS/G430L (blue), G750L (red) and IRAC 4.5\,$\mu$m and 3.6$\,\mu$m (magenta) transit observations, respectively. The gaps in seasons 2 and 3 are due to the monsoon season in southern Arizona where the observatory is located.}
\label{fig:henry_plot}
\end{figure}





We obtained nightly photometry of \wp~to monitor and characterise the stellar activity over the past three observing seasons with the Tennessee State University Celestron 14-inch (C14) automated imaging telescope (AIT) located at Fairborn Observatory in Arizona \citep[][]{henry99}.  The AIT uses an SBIG STL-1001E CCD camera and exposes through a Cousins R filter.  Each nightly observation of \wp~consists of 4--10 consecutive exposures that include several comparison stars in the same field of view.  The individual nightly frames were co-added and reduced to differential magnitudes (i.e. \wp~ minus the mean brightness of nine constant comparison stars).  Each nightly observation has been corrected for bias, flat-fielding, pier-side offset, and for differential atmospheric extinction.

A total of 258 nightly observations (excluding a few isolated transit observations) were acquired in five groups during three observing seasons between 2011 September and 2014 January. The five groups are plotted in Figure~\ref{fig:henry_plot}, where groups 2 through 5 have been normalized to have the same mean magnitude as the first group. The standard deviations of the five groups with respect to their corresponding means are 0.0032, 0.0065, 0.0034, 0.0078, and 0.0053 mag, respectively.  The precision of a single measurement in good photometric conditions with C14 is typically 0.002 - 0.003 mag (see, e.g., \citealt{sing13}).  The standard deviations of groups 2, 4, and 5 significantly exceed the measurement precision and thus indicate low-level activity due to star spots.  A periodogram analysis has been performed of each data set with trial frequencies ranging between 0.005 and 0.95 c/d, corresponding to a period range between 1 and 200 days. Period analysis of group five gives the clearest indication of rotational modulation with a period of $23.6\pm 0.5$~days and a peak-to-peak amplitude of 0.01 mag. We take this period to be our best determination of the star's rotation period.   While we do not rigorously standardize the individual differential magnitudes, the means of the five data groups before normalization appeared to vary from year by 0.005 to 0.01 mag.  We note that data groups 4 and 5, two of the three most active groups, were acquired after the $HST$ and $Spitzer$ observations.

The effect of stellar activity on the transmission spectrum can be taken into account by computing corrections of the measured $\rps$ provided the flux level of the star is known for each epoch of interest, as detailed in \cite{sing11b} and \cite{huitson13}. Unfortunately, our \hst~and \sp~observations occurred at epochs when our ground-based photometry was thin. However, it enables an evaluation of the uncertainties associated with the effect of stellar activity. Both the photometry in the relevant season, and the amplitude of the variation in previous seasons, suggest that the star flux varies by not more than about $1\%$ peak to peak or an r.m.s. of $0.3\%$ as measured from the ground-based light curve. Therefore, we choose to translate this value into an added uncertainty on the relative planet to star radius ratio (with a typical value of $\Delta \rps\simeq0.00022$) when presenting the transmission spectrum from all instruments following Equation 5 of \cite{sing11b}.

\begin{figure*}
\centering
\includegraphics[ trim = 4.5mm 1mm 8mm 0mm, clip, width=\textwidth,height=\textheight,keepaspectratio]{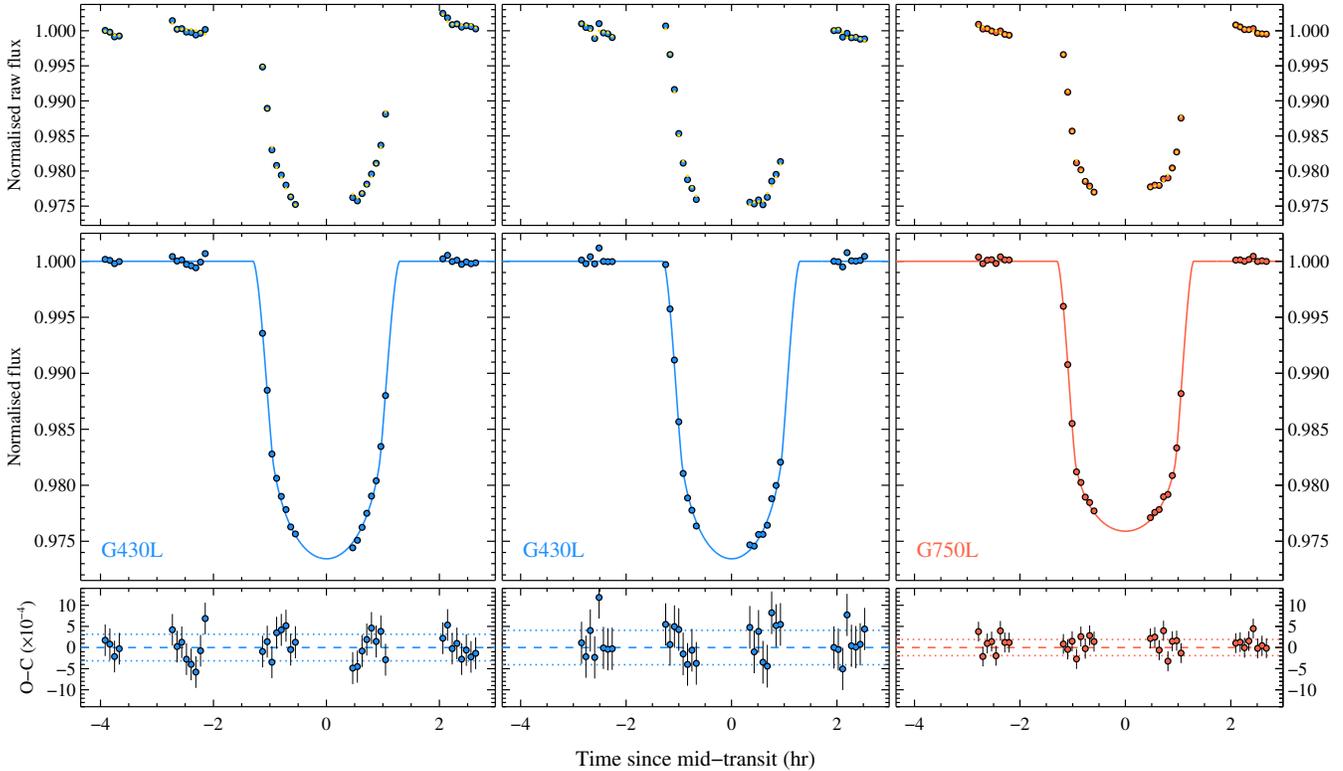}
\caption{ {\it{HST}}~STIS normalized white-light curves based on visits 9, 10 and 21 (left to right). Blue and red symbols indicate G430L and G750L gratings, respectively. Top panels: normalized raw flux with the best-fit model (yellow dots); Middle panels: detrended light curves along with the best-fit transit model \protect\citep{mandel02} plotted with continuous lines; Lower panels: light curve residuals and a $1\text{-}\sigma$ level (dotted lines).}
\label{fig:wlc_stis}
\end{figure*}


\section{Analysis}\label{sec:analysis}
We adopt similar analysis methods for the whole \hst~survey that are similar to the approach detailed in \cite{sing11b, sing13}, \cite{huitson13} and \cite{nikolov14}, which we also describe briefly here. We fit each \hst~and \sp~transit light curve in flux employing a two-component model that consists of a transit model multiplied by a systematics model. The transit model is based on the complete analytic formula given in \cite{mandel02}, which in addition to the central transit times ($T_{{\rm{C}}}$) and orbital period ($P$), is a function of the orbital inclination ($i$), normalized planet semi-major axis ($a/R_{\ast}$) and planet to star radius ratio ($R_{{\rm{p}}}/R_{\ast}$). The systematics models are different for the \hst~and \sp~data and are detailed below. We initially fixed the orbital period of the planet to its literature value \citep{gillon09a} before being updated to the final value reported in Section\,\ref{sec:transit_ephemeris}. 

\subsubsection{\stis white light curve modelling}\label{sec:hst_analysis}
The three \stis white-light curves were fit simultaneously with common inclination, semi-major axis and a planet-to-star radius contrast. We internally linked the radius parameter in our code to produce joint value for both G430L gratings. 
For the white-light curves we fit these parameters simultaneously assuming flat priors. We use the resulting parameters from both G430L and G750L white-light curves in conjunction with literature results to refine \wp's system parameters and transit ephemeris.

 We take into account the stellar limb-darkening of the parent star in our code adopting the four parameter non-linear limb-darkening law. We calculate the relevant coefficients with {\tt{ATLAS}} stellar models following \cite{sing10} and adopting the stellar  parameters $T_{{\rm{eff}}} = 5375$\,K, $\log{g}=4.61$ and [Fe/H]=-0.2 from \cite{gillon09a} and \cite{doyle13}. We choose to rely on theoretically derived stellar limb-darkening coefficients from {\tt{1D}} stellar models, rather than to fit for them in the data in order to reduce the number of free parameters in the fit. Furthermore, this approach also eliminates the well-known wavelength-dependent degeneracy of limb darkening with transit depth \citep{sing08a}.
 
 \begin{figure}
\begin{subfigure}
  \centering
  \includegraphics[trim={0 1.35cm 0 0}, clip=true, width=0.5\textwidth]{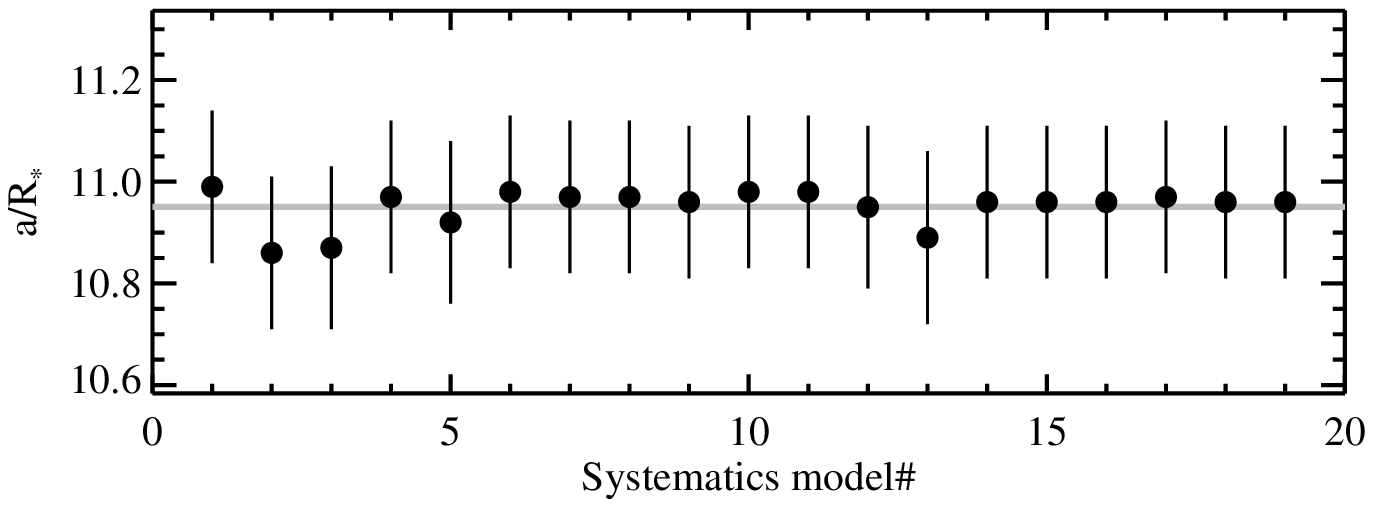}
\end{subfigure}
\begin{subfigure}
  \centering
  \includegraphics[trim={0 1.35cm 0 0.6cm}, clip=true, width=0.5\textwidth]{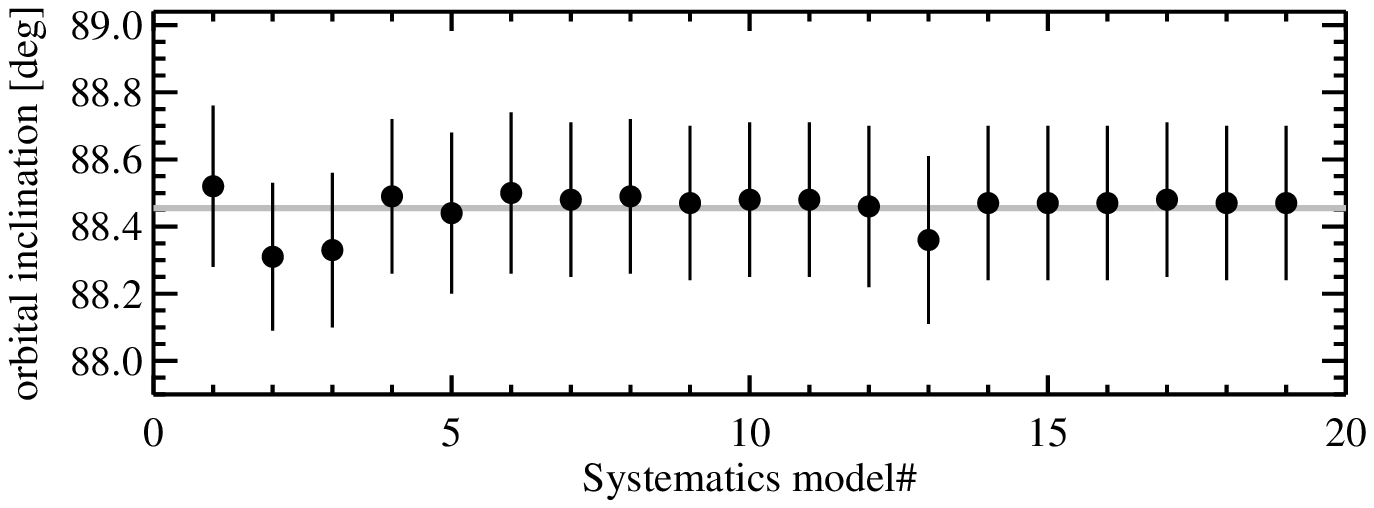}
\end{subfigure}
\begin{subfigure}
  \centering
  \includegraphics[trim={0 1.35cm 0 0.6cm}, clip=true, width=0.5\textwidth]{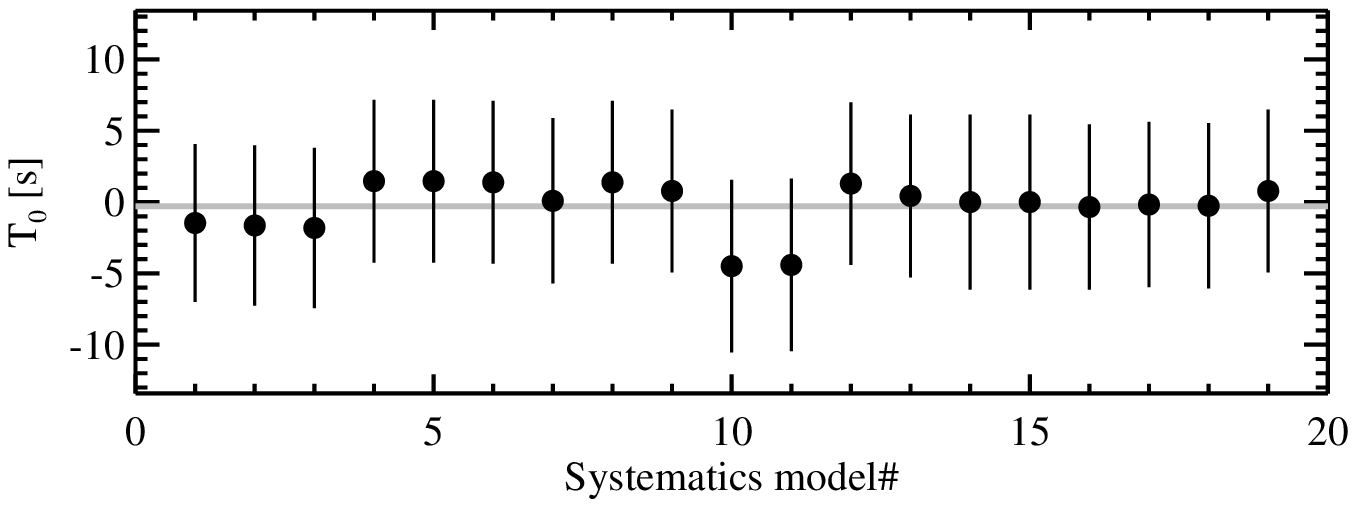}
\end{subfigure}
\begin{subfigure}
  \centering
  \includegraphics[trim={0 0 0  0.6cm}, clip=true, width=0.5\textwidth]{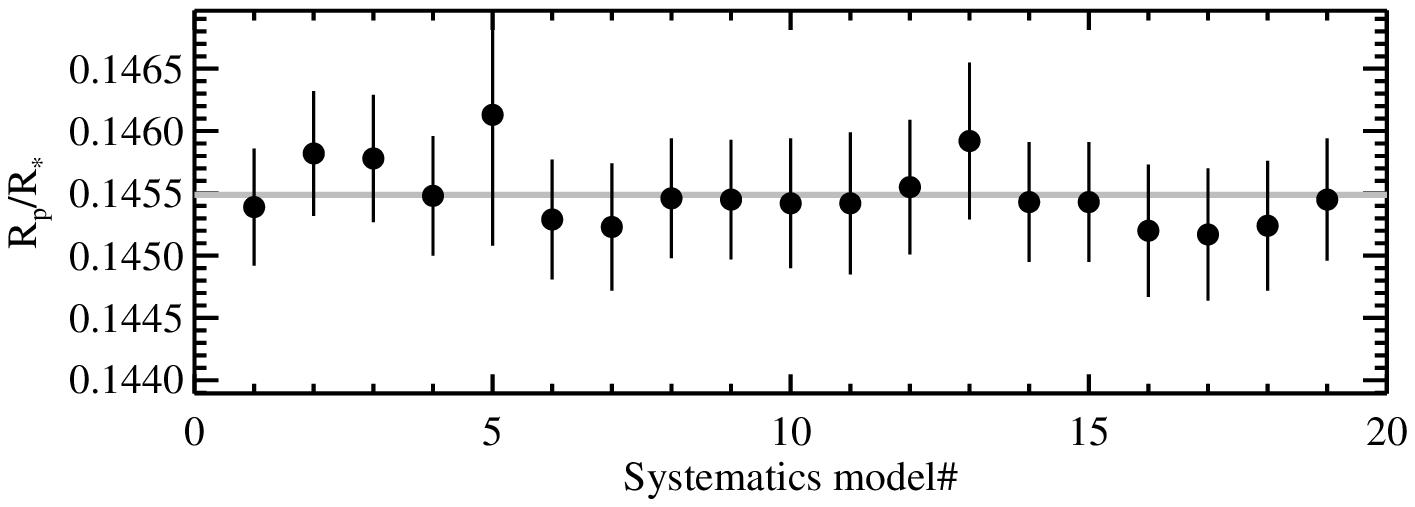}
\end{subfigure}
\caption{Transit parameters as a function of systematics model for \stis G750L data with  uncertainties derived assuming photon noise only (i.e. not rescaled). The horizontal axis indicates the systematics model identification number as follows:\\
1)~$\phi_{t} + \phi^{2}_{t}+ \phi^{3}_{t}+ \phi^{4}_{t}$ \\
2)~$\phi_{t} + \phi^{2}_{t}+ \phi^{3}_{t}+ \phi^{4}_{t}+ \phi^{5}_{t}$ \\
3)~$\phi_{t} + \phi^{2}_{t}+ \phi^{3}_{t}+ \phi^{4}_{t}+ \phi^{5}_{t}+ \phi^{6}_{t}$ \\
4)~$\phi_{t} + \phi^{2}_{t}+ \phi^{3}_{t}+ \phi^{4}_{t} + t$\\
5)~$\phi_{t} + \phi^{2}_{t}+ \phi^{3}_{t}+ \phi^{4}_{t} + t + t^2$\\
6)~$\phi_{t} + \phi^{2}_{t}+ \phi^{3}_{t}+ \phi^{4}_{t} + t^2$\\
7)~$\phi_{t} + \phi^{2}_{t}+ \phi^{3}_{t}+ \phi^{4}_{t} + t + x$\\
8)~$\phi_{t} + \phi^{2}_{t}+ \phi^{3}_{t}+ \phi^{4}_{t} + t + y$\\
9)~$\phi_{t} + \phi^{2}_{t}+ \phi^{3}_{t}+ \phi^{4}_{t} + t + \omega$\\
10~$\phi_{t} + \phi^{2}_{t}+ \phi^{3}_{t}+ \phi^{4}_{t} + t + x+x^2$\\
11)~$\phi_{t} + \phi^{2}_{t}+ \phi^{3}_{t}+ \phi^{4}_{t} + t + x + x^2 + x^3$\\
12)~$\phi_{t} + \phi^{2}_{t}+ \phi^{3}_{t}+ \phi^{4}_{t} + t + y + y^2 $\\
13)~$\phi_{t} + \phi^{2}_{t}+ \phi^{3}_{t}+ \phi^{4}_{t} + t + x + y^2 + y^3$\\
14)~$\phi_{t} + \phi^{2}_{t}+ \phi^{3}_{t}+ \phi^{4}_{t} + t + \omega+ \omega^2$\\
15)~$\phi_{t} + \phi^{2}_{t}+ \phi^{3}_{t}+ \phi^{4}_{t} + t + x + x^2 + x^3$\\
16)~$\phi_{t} + \phi^{2}_{t}+ \phi^{3}_{t}+ \phi^{4}_{t} + t + x + y + \omega$\\
17)~$\phi_{t} + \phi^{2}_{t}+ \phi^{3}_{t}+ \phi^{4}_{t} + t + x + x^2 + x^3$\\
18)~$\phi_{t} + \phi^{2}_{t}+ \phi^{3}_{t}+ \phi^{4}_{t} + t + x +\omega$\\
19)~$\phi_{t} + \phi^{2}_{t}+ \phi^{3}_{t}+ \phi^{4}_{t} + t + y + \omega$\\
}
\label{fig:tpar_model}
\end{figure}

\begin{table*}
\centering
\caption{1D limb-darkening coefficients employed in the fit to the \stis white light curves and results for the \rps, transit times and residual scatter in parts-per-million (ppm).}
\begin{tabular}{@{} c c c c c c c c}
\hline
\hline
Visit ID   &  Instrument  &   $c_{1}$  &   $c_{2}$    &    $c_{3}$    &    $c_{4}$  &   \rps  & r.m.s. (ppm)\\
\hline
9 & STIS/G430L & 0.6246 & -0.4673 & 1.2881 & -0.6067 & $0.14726 \pm 0.00086$ &  318\\
10 & STIS/G430L & 0.6246 & -0.4673 & 1.2881 & -0.6067 & $0.14726 \pm 0.00086$ &  372\\
21 & STIS/G750L & 0.6991 & -0.5421 & 1.0896 & -0.5187 & $0.14520 \pm 0.00061$ &  191\\
\hline
\end{tabular}
\label{tab:wlc}
\end{table*}

\begin{figure*}
\centering
\includegraphics[scale=0.25,keepaspectratio]{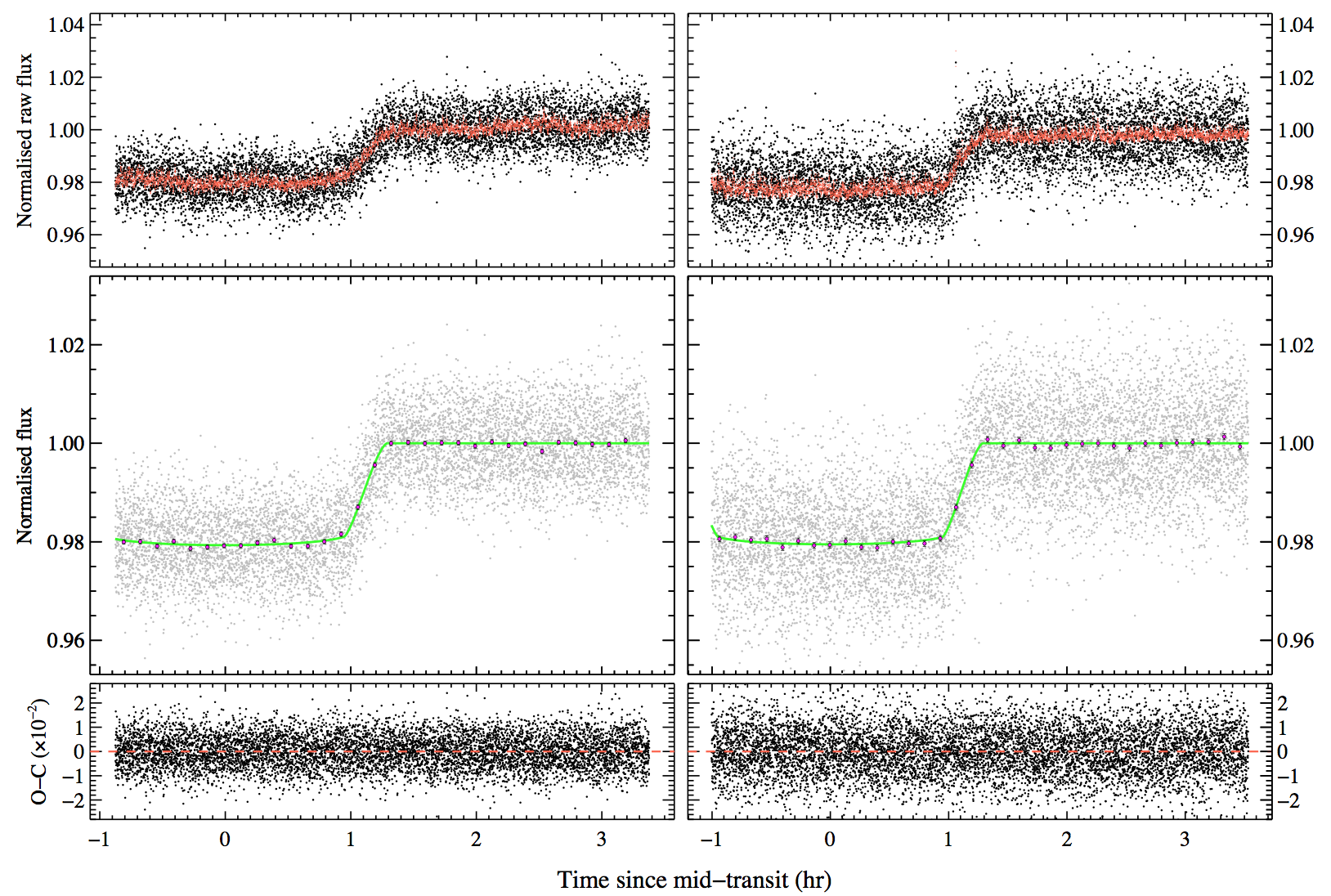}
\caption{Warm \sp~\ir~$3.6$~and $4.5~\mu$m photometry (left and right respectively). Top panels: Raw flux and the best-fit transit and systematics model; Middle panels: Detrended light curves and the best-fit transit models and binned by 8 minutes; Lower panels: Light curve residuals.}
\label{fig:spitzer_data}
\end{figure*}

Orbit-to-orbit flux corrections were applied to the \stis data similar to past \hst~studies by fitting a fourth-order polynomial of the spacecraft orbital phase ($\phi_{t}$).  In addition, we find that the systematics vary from orbit-to-orbit (i.e. dependence with time, $t$), which is a known effect from previous {\stis} studies as well as with the detector positions of the spectra, as determined by the spectral trace orientation $(x,y)$ obtained  from the {\tt{APPAL}} task in {\tt{IRAF}} and the wavelength shift ($\omega$) of each spectrum of the time series, compared to a reference. Model selection was investigated as in \cite{nikolov14} including  i) polynomials of degrees higher than fourth-order for the {\sl{HST}} orbital phase, and ii) additional terms to the planet orbital phase, spectral shifts and traces. Such models were found statistically unjustified based on the Bayesian Information Criterion (BIC; \citealt{schwarz78}). For instance, we found that fifth or sixth order polynomials of \hst~orbital phase gave BICs of 149 and 153, respectively. These values are much higher compared to the value produced by the most favourable model which includes a fourth order polynomial, a linear time term and  $x$ (BIC of 96). The final systematics model for the red data include a fourth order polynomial of \hst~orbital phase, a linear time term, and $\omega$ with BIC of 77. For comparison a systematics model of a fourth-order polynomial of \hst~orbital phase, a linear time term, $\omega$ and $\omega^2$ gave BIC of 83. Finally, we find that the fitted transit parameters are very well-constrained (including the transit depth) and are consistent regardless of the exact systematics model (see Fig.~\ref{fig:tpar_model}).

The errors on each photometric data point from the \hst~time series were initially set to the pipeline values, which are dominated by photon noise with readout noise also taken into account. We determine the best-fitting parameters simultaneously with the Levenberg-Marquardt least-squares algorithm as implemented in the {\tt{IDL\footnote{The acronym {\tt{IDL}} stands for Interactive Data Language.} {\tt{MPFIT}}}}\footnote{\url{http://www.physics.wisc.edu/~craigm/idl/fitting.html}} package \citep{markwardt09} using the unbinned data. The final results for the uncertainties of the fitted parameters were taken from MPFIT after we rescaled the errors per data point based on the standard deviation of the residuals. We compare the residual dispersion of each white light curve to the expected one from photon noise. The data is found to reach $\sim77, \sim65$ and $\sim89\%$ of the photon noise limit for visits 9, 10 and 21, respectively which is similar earlier STIS studies \citep{huitson13, sing13, nikolov14}.

Finally, we also explored the dependence of the fitted transit parameters of the adopted systematics model. In these fits we report the fitted parameters with uncertainties assuming pure photon noise and no rescaling. A summary of our results for grating G750L is shown in Fig~\ref{fig:tpar_model}. We find that the fitted parameters in each \stis grating remain practically the same regardless the systematics model. These results are similar to what was found in \cite{nikolov14}.
\subsubsection{\sp~\ir~light curve fits}\label{sec:irac_analysis}
We model the $3.6$ and $4.5~\mu$m transit light curves following standard procedures for warm \sp. In particular we correct the intrapixel sensitivity induced flux variations by fitting a polynomial function of the stellar centroid position \citep{reach05, charbonneau05, charbonneau08, knutson08}. As discussed in \cite{lewis13} this method has been proved to work reasonably well on short timescales ($<10$ hr) where the variations in the stellar centroid position are small ($<0.2$ pixels). These two conditions are satisfied for the \wp~data analysed in our study. To correct for systematic effects, we used a parametric model  of the form: $f(t) = a_{0} + a_{1}x  + a_{2}x^2  + a_{3}y  + a_{4}y^2  + a_{5}xy + a_{6}t$, where $f(t)$ is the stellar flux as function of time, $x$ and $y$ are the positions of the stellar centroid on the detector, $t$ is time and $a_{0}$ to $a_{6}$ are the free parameters of the fit. The limb-darkening of the star was taken into account by a non-linear law, assuming an ATLAS stellar atmosphere and the recipe of \cite{sing10}. Both light curves were jointly fitted with all parameters set free except to the orbital inclination and semimajor axis (as \ar). These two parameters were linked to result to single values. The best-fit orbital parameters were found to be in excellent agreement (within $1\text{-}\sigma$) with the results from the \stis white-light curve analysis. We estimated the degree of red noise using the \cite{carter09} wavelet method finding white and red noise components $\sigma_{w}=0.0068$ and $\sigma_{r}=0.00012$ for the $3.6~\mu$m data and $\sigma_{w}=0.0094$ and $\sigma_{r}=0.00016$ for the $4.5~\mu$m light curve, respectively. The values for \rps~used for the transmission spectrum were measured by fixing the system parameters, namely the orbital period, \ar, $i$ and the central transit times to their best-fit values and the limb-darkening coefficients were kept fixed to their theoretical values. 

We also performed two tests to explore the impact of incomplete \sp~\ir~transit light curves on the measured system parameters. We obtained the $3.6$ and $4.5~\mu$m full-transit coverage light curves of WASP-31 reported in Sing et al. 2014. The data has similar noise properties and has been reduced with the same pipeline as our \wp~data. The WASP-31 light curves conveniently include enough out-of-transit baseline on both ends of the transits to evaluate measuring full vs half transits. We trimmed $20\times n{\rm{-minute}}$ sections ($n=1,7$)  from the beginning of the observation to produce half-transit light curves until we pass the transit central time. We fitted each light curve following the same methodology adopted for the \wp~\ir\,data. In the first test, we fitted for all transit and systematics parameters while in the second we fixed \ar, the orbital inclination and the central times to their best-fit values measured using the whole data sets. We find that the first test gave transit parameters consistent with each other within $1\text{-}\sigma$ when using light curves containing the transit central time. The largest variation is found for the values of transit mid-time and \ar. The radius values (i.e. \rps) for both channels remained within one sigma with increasing errors due to the smaller light curve sections used. We measure r.m.s. of \rps\,for both channels to be   0.0025 and 0.0016, respectively. The second test aimed to explore the reliability of the measured \rps\,from half transits. We find this parameter more stable than in the first test with r.m.s. of \rps\,values of 0.0013 and 0.0010 for the  $3.6$ and $4.5~\mu$m light curves, respectively. We conclude that the \rps\,values measured with our fitting code from incomplete \sp~\ir~$3.6$ and $4.5~\mu$m can be trusted as long as the data includes the transit mid-time and the fitting is done by fixing the system parameters and limb-darkening coefficients to well determined values. As the WASP-6 light curves are only missing pre-ingress data the obtained planet radii should be reliable. 



     
\subsection{System parameters and transit ephemeris}\label{sec:transit_ephemeris}

\begin{figure}
        \centering
                 \includegraphics[trim = 7.5mm 3.5mm 3mm 5mm, clip, scale=0.75]{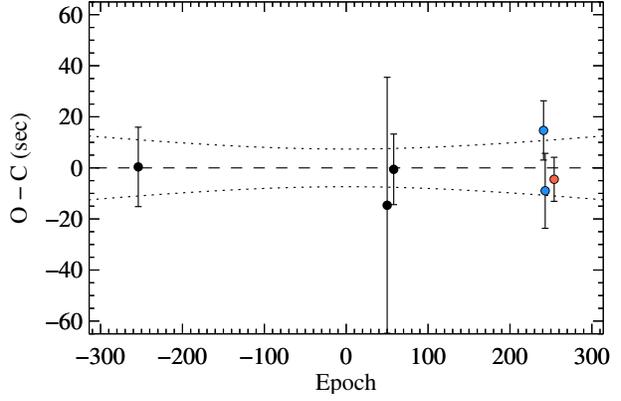}
         \caption{Observed minus computed (O-C) transit times for \wp b, based on our \hst~\stis measurements (blue and red dots) and observations reported in the literature (black dots) and the best-fit orbital period and central transit time derived from our analysis. The dotted lines represent the computed $1\text{-}\sigma$ uncertainty of the ephemeris.}
\label{fig:varmon}
\end{figure}

We combined the central transit times from our \hst\, dataset with the transit times reported by \cite{dragomir11} and  \cite{Jordan13} to derive an updated transit ephemeris of \wp\,b. In addition, we obtained the archival light curve with full transit coverage reported in \cite{gillon09a}\footnote{http://exoplanetarchive.ipac.caltech.edu/starsearch.html} and performed a fit to determine its central time. We excluded the two \sp~\ir~light curves as these originate from incomplete transits and are not useful for determination of accurate transit ephemeris. We fitted the remaining measurements presented in Table\,\ref{tab:octab} and Fig.\,\ref{fig:varmon}  with a linear function of the orbital period ($P$) and transit epoch ($E$),
\begin{equation}
  T_{\mathrm{{{\rm{C}}}}}(E) = T_{\mathrm{0}} + E P.
\end{equation}       
We find a period of $P =  3.36100239 \pm 0.00000037$ (day) and a mid-transit time of $ T_{\mathrm{{{\rm{C}}}}} = 2455278.716260 \pm     0.000085$ (day).

In addition, we refined the system parameters including the orbital inclination, $i$ and normalized semimajor axis, (\ar). We performed a joint fit to the three \hst~white-light curves finding $i=88.57\pm 0.29^{\circ}$ and \ar$ = 11.05 \pm 0.17$.



\begin{table}
\centering
\caption{Transit mid-times and O-C residuals computed using our \hst~\stis light curves and results from the literature.}
\begin{tabular}{@{} r c c c }
\hline
\hline
Epoch & Central time & O$-$C & Reference\\
 & (${\rm{BJD}}_{{\rm{TDB}}}$) & (days) & \\
\hline
 -254 & $24  54425.021657\pm{ 0.000180}$ &   0.000004 &        1\\
   50 & $24  55446.766210\pm{ 0.000580}$ &  -0.000169 &        2\\
   58 & $24  55473.654392\pm{ 0.000160}$ &  -0.000007 &        3\\
  241 & $24  56088.718006\pm{ 0.000134}$ &   0.000170 &        4\\
  243 & $24  56095.439737\pm{ 0.000170}$ &  -0.000104 &        4\\
  254 & $24  56132.410816\pm{ 0.000100}$ &  -0.000052 &        4\\
\hline
\multicolumn{4}{l}{Reference: $1-$\cite{gillon09a}; $2-$\cite{dragomir11}; }\\
\multicolumn{4}{l}{$3-$\cite{Jordan13}, $4-$this work. }
\end{tabular}
\label{tab:octab}
\end{table}

%
\begin{figure*}
\centering
\includegraphics[width=0.8\textwidth,height=\textheight,keepaspectratio]{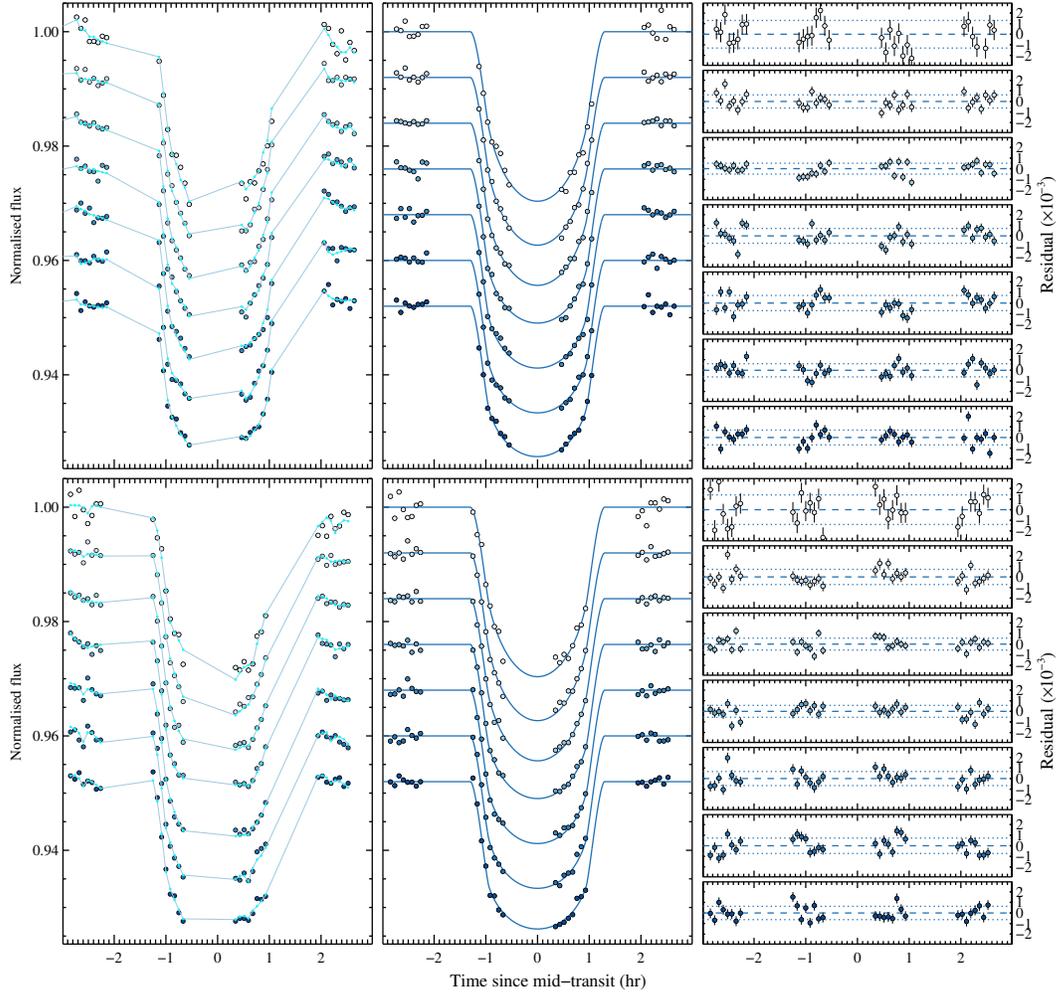}
\caption{\protect{{\hst}~{\stis}} G430L spectral bin light curves of visits 9 and 10. Left panels: raw light curves and the best-fit transit model, multiplied in flux by a systematics model shifted with an arbitrary constant. The points have been connected with lines for clarity and the light curves are presented in wavelength with the shortest wavelength bin displayed at the top and longest wavelength bin at the bottom. Middle panels: corrected light curves and the best-fit transit model. Right panels: observed minus computed residuals with 1-$\text{-}\sigma$ error bars along with the standard deviation (dotted lines).}
\label{fig:bluelcfig}
\end{figure*}

\begin{figure*}
\centering
\includegraphics[width=0.85\textwidth,height=\textheight,keepaspectratio]{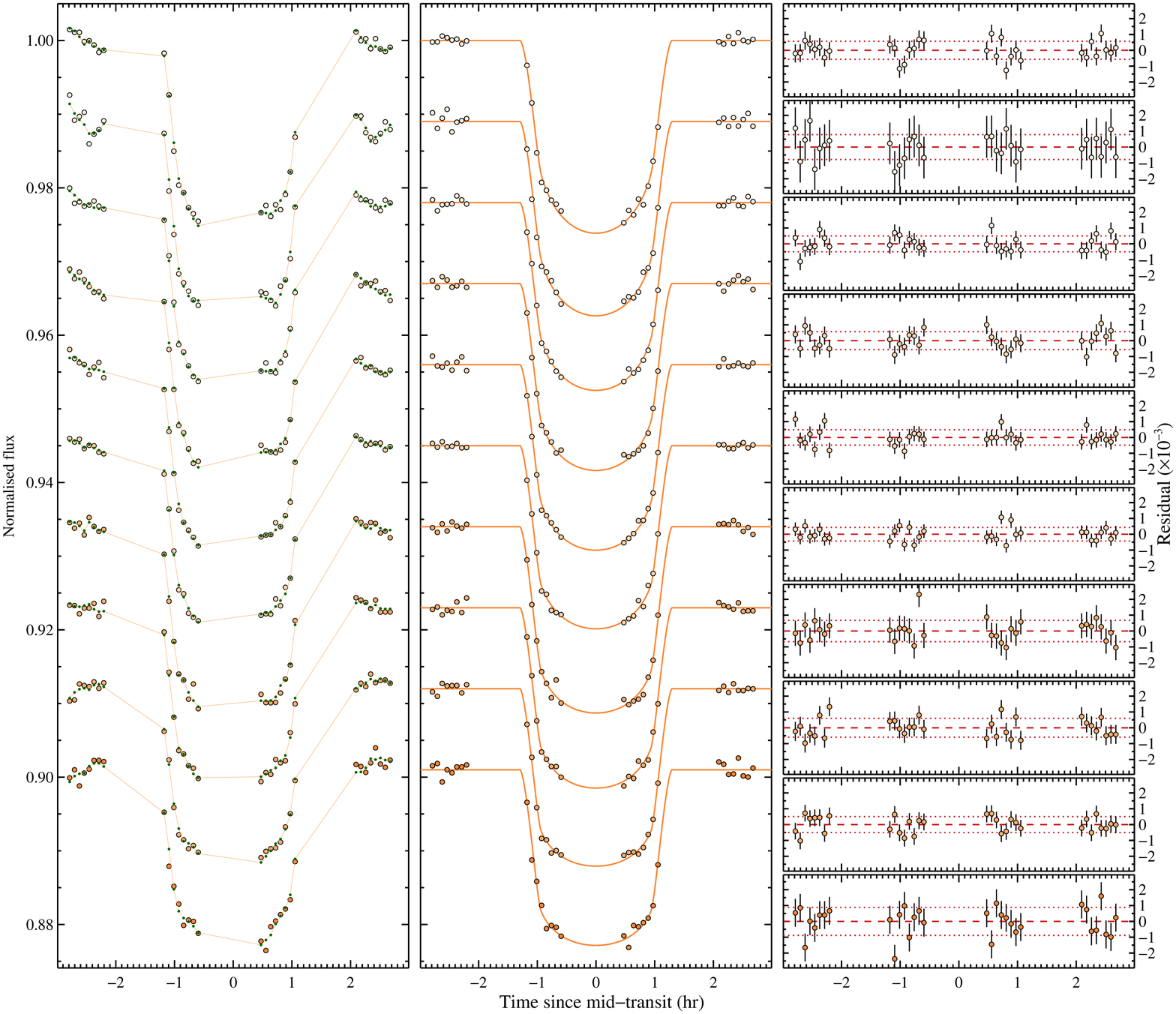}
\caption{Same as Fig.~\ref{fig:bluelcfig}, but for the {\it{HST}}~STIS G750L data (Visit 21).}
\label{fig:redlcfig}
\end{figure*}

\subsection{An evaluation of the stellar activity from the \hst~data}\label{sec:transit_ephemeris}
\begin{figure}
\centering
\includegraphics[trim = 5mm 3.5mm 3mm 5mm, clip, scale=0.77]{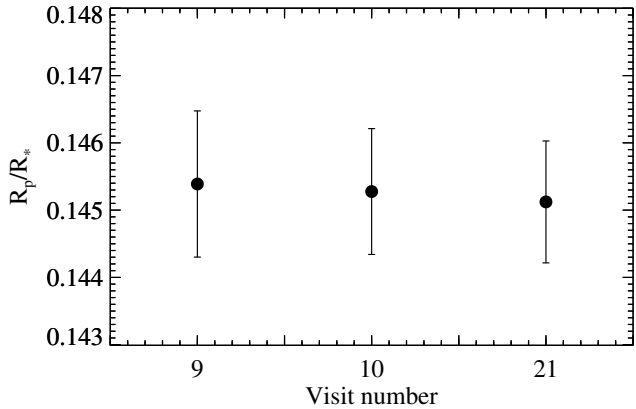}
\caption{The three measured transit radii in a bin between $5250-5400$~\AA~taken between the two \stis G430L visits and the \stis G750L visit. No significant difference in the transit depth is seen.}
\label{fig:hst_activity}
\end{figure}

The overlapping region between the two \stis~gratings offers the opportunity to measure the photometric variability of the host star over the time interval of 40 days covered by the three \hst~visits, as significant stellar activity from stellar spots would change the measured transit radii. We used three light curves from the range $5400-5650$~\AA~(i.e. avoiding the rapid sensitivity change in G750L from $5250-5400$~\AA, see Fig. 1 from \citealt{nikolov14}) and measured the quantity $\rps$ from a fit, keeping the system parameters, ephemeris and limb darkening to their best-fit white light values and theoretical coefficients, respectively. The result is shown in Fig.\ref{fig:hst_activity}. The standard deviation in $\rps$ is $\sim2\times10^{-4}$ and the translated flux variation of the star is $<0.3\%$, which is in agreement with our ground-based result discussed in section~\ref{sec:stellaractivity} and the result of \cite{Jordan13} who found a peak-to-peak amplitude $<4$ mmag. Our result suggests that the three \hst~visits occurred at similar stellar flux levels implying no significant offset between the blue and red \stis~sections of the transmission spectrum. Finally, the very similar radius values for the two blue visits obtained six days apart imply the lack of variability related with the planetary atmosphere itself.

\subsection{Fits to transmission spectra}

The primary science goal of this study is to construct a low-resolution optical to near-infrared transmission spectrum of \wp\,b and to pursue the prediction of strong optical absorbers such as sodium (observed in the Na\,{\sc i} doublet at $\lambda=5893$\,\AA), potassium (K\,{\sc{i}} doublet at $\lambda=7684$\,\AA) and water vapour (e.g. in $1.45~\mu$m band) or alternatively high-altitude atmospheric haze (in optical scattering), and/or a constant flat scattering at $\lambda \gtrsim 1~\mu$m. Light curves from various spectral bins were extracted from the STIS/G430L and G750L data to construct a broad-band transmission spectrum and to search for expected strong absorbers (see Fig. \ref{fig:bluelcfig} and  \ref{fig:redlcfig}).

As pointed out in \cite{pont13} three issues must be addressed in order to place radius measurements on a common scale and build a transmission spectrum spanning a wide wavelength coverage: common orbital parameters, stellar limb darkening and the effect of stellar variability and star spots. We take into account all of these issues in the analysis of the transmission spectrum from the \hst~and \sp~data. In the light curve fits we fixed the system parameters, i.e. the orbital period ($P$), inclination ($i$), normalized semi-major axis ($a/R_{\ast}$) and the transit mid-times ($T_{\rm{C}}$) to their values derived from the white-light curve analysis (Section~\ref{sec:transit_ephemeris}). We fit for the relative radius of the planet and the parameters describing the instrumental systematics.  The activity of the parent star was taken into account by inflating the error bars of the measured \rps of each spectral bin assuming stellar variability of $0.3\%$. This number also takes into account the contribution of unocculted spots. The four non-linear limb-darkening coefficients were fixed to their theoretical values, computed from stellar models and taking into account the instrument response. 

\begin{figure}
\centering
\includegraphics[trim = 0 0 0 0, clip, width = 0.5\textwidth]{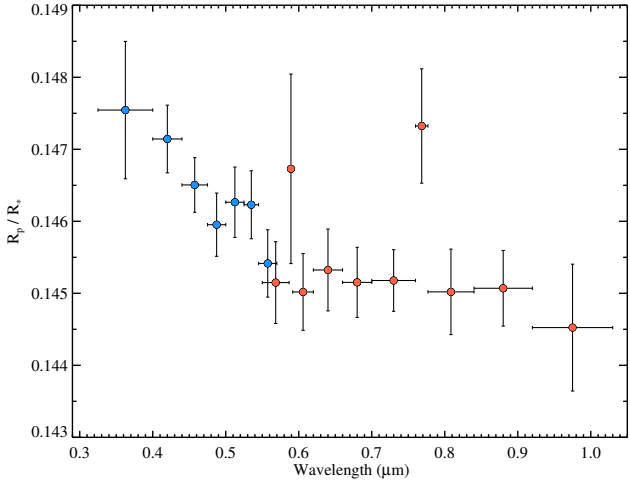}
\caption{ Combined \hst~\stis\,G430L and G750L transmission spectrum of \wp b (blue and red dots, respectively).}
\label{fig:stis_spectrum}
\end{figure}

Similar to \cite{sing13} and  \cite{nikolov14} we performed the modelling of systematic errors in each spectral bin by two methods. In the first method, we fit each light curve independently with a parametrized model. In the second approach, we removed the common-mode systematics from each spectral bin before fitting for residual trends with a parametrized model but with fewer parameters. The common-mode trends were extracted at the white-light curve analysis by dividing the white-light raw flux in each grating to the best-fit transit model constrained from a joint fit to the three \stis~data sets. It was found that this method reduces the amplitude of the breathing systematics. We found that for the \wp~data the common-mode approach generally worked better for the G430L data than for the G750L spectral light curves. This is expected as the G430L spectral light curves exhibit very similar systematic trends which is not the case for the G750L spectral light curves. A visual inspection of the out-of-transit data on Fig.\ref{fig:bluelcfig} and  \ref{fig:redlcfig} reveals this fact. As found in \cite{sing13} for the WASP-12b and \cite{nikolov14} for the HAT-P-1b \stis~transit data, we find that the transmission spectra from both methods show little difference (i.e. deviations in the \rps within $1\text{-}\sigma$ of their fitted values). For consistency we choose to report the final transmission spectrum from the first method. The broad-band spectrum results are provided in Table~\ref{tab:specfit_rad} and displayed in Fig.\ref{fig:stis_spectrum}.



\begin{table*}
\centering
\caption{Measured $R_{\bf{p}}/R_{\ast}$ from fits to the G430L and G750L transit light curves and limb darkening coefficients.}
\begin{tabular}{@{} c c c c c c}
\hline
\hline
Wavelength (\AA)   &  \rps     &  $c_{1}$  &   $c_{2}$    &    $c_{3}$    &    $c_{4}$\\
\hline
Visit        9 \&       10 & & & & & \\
$3250 - 4000$ & $0.14755 \pm 0.00098$ & 0.5049 & -0.6907 & 1.9020 & -0.7850\\
$4000 - 4400$ & $0.14714 \pm 0.00052$ & 0.5077 & -0.6398 & 1.8008 & -0.7416\\
$4400 - 4750$ & $0.14650 \pm 0.00044$ & 0.6469 & -0.8914 & 1.9172 & -0.7695\\
$4750 - 5000$ & $0.14595 \pm 0.00049$ & 0.5652 & -0.3541 & 1.2492 & -0.5989\\
$5000 - 5250$ & $0.14627 \pm 0.00054$ & 0.5392 & -0.2004 & 1.0534 & -0.5399\\
$5250 - 5450$ & $0.14623 \pm 0.00052$ & 0.6247 & -0.4675 & 1.2882 & -0.6067\\
$5450 - 5700$ & $0.14541 \pm 0.00052$ & 0.5404 & -0.2277 & 1.0662 & -0.5468\\
Visit       21  & & & & & \\
$5500 - 5868$ & $0.14515 \pm 0.00061$ & 0.6081 & -0.4022 & 1.2199 & -0.5927\\
$5868 - 5918$ & $0.1467 \pm 0.0013$ & 0.5750 & -0.2207 & 0.9647 & -0.5064\\    
$5918 - 6200$ & $0.14502 \pm 0.00058$ & 0.6241 & -0.3386 & 1.0136 & -0.5079\\
$6200 - 6600$ & $0.14532 \pm 0.00061$ & 0.6552 & -0.4611 & 1.1582 & -0.5801\\
$6600 - 7000$ & $0.14515 \pm 0.00053$ & 0.6791 & -0.4802 & 1.0974 & -0.5326\\
$7000 - 7599$ & $0.14518 \pm 0.00048$ & 0.7361 & -0.6490 & 1.2003 & -0.5554\\
$7599 - 7769$ & $0.14732 \pm 0.00082$ & 0.7099 & -0.5449 & 1.0219 & -0.4847\\
$7769 - 8400$ & $0.14502 \pm 0.00063$ & 0.7578 & -0.7132 & 1.2070 & -0.5521\\
$8400 - 9200$ & $0.14507 \pm 0.00057$ & 0.7005 & -0.5541 & 0.9344 & -0.4345\\
$9200 - 10300$ & $0.14452 \pm 0.00091$ & 0.7221 & -0.6191 & 0.9967 & -0.4582\\

$36000$            &   $ 0.1404 \pm 0.0014$ &   0.4434  &  -0.2253   &    0.1829    & -0.0717\\
$45000$            &   $ 0.1405 \pm 0.0015$ &   0.5356   &  -0.6254  &    0.6113  &   -0.2243\\
\hline
\end{tabular}
\label{tab:specfit_rad}
\end{table*}


\section{Discussion}\label{sec:discussionsec}

\subsection{A search for narrow-band spectral signatures}\label{sec:narrow_sign_sec}

We performed a thorough search for narrow-band absorption features including the expected signatures of Na, K, H$_{\alpha}$ and H$_{\beta}$. In the context of the \hst~survey, sodium has been detected by \cite{nikolov14} in the HAT-P-1b exoplanet using a $30$\,\AA~bin and potassium has been detected in the WASP-31b atmosphere by Sing et al. (2014) using 75\,\AA\,bin. In the case of \wp b, we followed the methods described in \cite{nikolov14} and used a set of six bands centred on the expected features with widths from 15 to 90\,\AA~with step of 15\,\AA~for Na, H$_\alpha$ and H$_\beta$ and twice larger for the potassium feature. We find only tentative evidence for sodium and potassium line cores with the largest significance levels of 1.2$\text{-}\sigma$ and 2.7$\text{-}\sigma$ in $50$ and $170$\,\AA~bins, respectively. We also searched but did not find evidence for Na and K pressure-broadened line wings. Finally, we find also no evidence for H$_{\alpha}$ nor H$_{\beta}$.

\subsection{Comparison to existing cloud-free theoretical models}\label{sec:narrow_sign_sec}

We compare our transmission spectrum to a variety of different cloud-free atmospheric models based on the formalism of \cite{burrows10}, \cite{howe12} and \cite{fortney08,fortney10}. We averaged the model within the transmission spectrum wavelength bins and fitted these theoretical values to the data with a single free parameter that controls their vertical position. We compute the $\chi^2$ statistic to quantify model selection with the number of degrees of freedom for each model given by  $\nu = N-m$, where $N$ is the number of data points (21) and $m$ is the number of fitted parameters.

The models from \cite{burrows10} and \cite{howe12} assume {\tt{1D}} dayside temperature~--~pressure \tp profile with stellar irradiation, in radiative, chemical, and hydrostatic equilibrium. Chemical mixing ratios and corresponding opacities assume solar metallicity and local thermodynamical chemical equilibrium accounting for condensation with no ionisation, using the opacity database from \cite{sharp07} and the equilibrium chemical abundances from \cite{burrows99} and \cite{burrows01}.

The models based on \cite{fortney08,fortney10} include a self-consistent treatment of radiative transfer and chemical equilibrium of neutral and ionic species. Chemical mixing ratios and opacities assume solar metallicity and local chemical equilibrium, accounting for condensation and thermal ionisation though no photochemistry \citep{lodders99, lodders02a,lodders02b, lodders06, visscher06, freedman08, lodders09}. In addition to isothermal models, transmission spectra were calculated using {\tt{1D}} temperature-pressure \tp profiles for the dayside, as well as an overall cooler planetary-averaged profile. Models were also generated both with and without the inclusion of TiO and VO opacities. Because each of the aforementioned models started at wavelength $\lambda = 3500$~\AA~(due to a restriction of the employed opacity database), we extrapolated the models in the range $2700-3500$~\AA~to enable self-content comparisons to the bluest \stis measurements along with all models.

\begin{figure}
\centering
\includegraphics[trim = 0 0 0 0, clip, width = 0.5\textwidth]{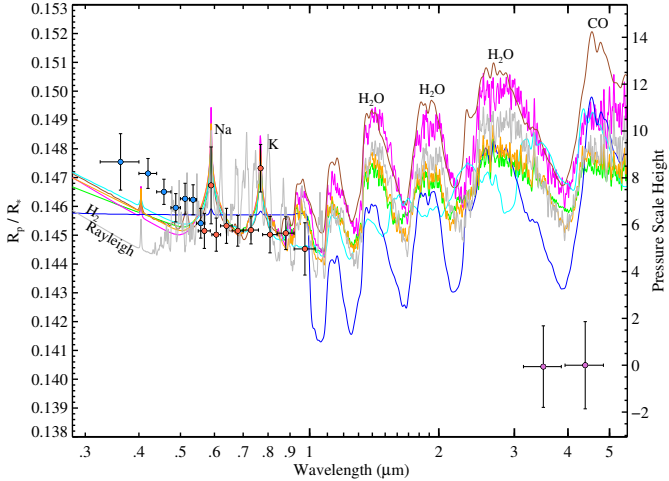}
\caption{ \wp b transmission spectrum compared to seven different cloud-free atmosphere models listed in Table\,\ref{tab:mod_stat1}, including Burrows-Dayside1200K-$3\times$solar (brown),  Burrows-Isoth.1200K-high-CO (cyan), Burrows-Isoth.1200K-ExtraAbsorber (blue), Fortney-Isoth.1000K (green), Fortney-Isoth.1500K-noTiO/VO (magenta), Fortney-PlanetAveraged1200K (orange) and Fortney-Isoth.1500K-withTiO/VO (grey). All these models hardly reproduce the two \sp~measurements. Prominent absorption features are labeled.}
\label{fig:cloud_free}
\end{figure}

As it becomes apparent from Table~\ref{tab:mod_stat1} all of the cloud-free models struggle to represent the observed  transmission spectrum of \wp b, providing poor fits (see Fig.~\ref{fig:cloud_free}).



\begin{table}
\centering
\caption{Model (cloud-free) selection fit statistics ($N=19$).}
\begin{tabular}{@{} l c c c c}
\hline
\hline
Models & $\chi^2$ & $\nu$  & BIC \\
\hline
Burrows-Isoth.1200K-ExtraAbsorber      &       55.9     &      17  &        58.9  \\
Fortney-Isoth.1000K                    &       74.1   &              18  &        77.0  \\
Fortney-PlanetAveraged1200K            &       81.9   &              18  &        84.8  \\
Burrows-Isoth.1200K-high-CO            &       82.7   &              17  &        85.6  \\
Burrows-Dayside1200K-$3\times$solar    &      102.9   &      18  &       105.9  \\
Fortney-Isoth.1500K-noTiO/VO           &      107.6   &             18  &       110.5  \\
Fortney-Isoth.1500K-withTiO/VO         &      124.4   &            18  &       127.4  \\
\hline	
\end{tabular}
\label{tab:mod_stat1}
\end{table}

\subsection{Interpreting the transmission spectrum with aerosols}\label{sec:narrow_sign_sec}

An overall slope in the transmission spectrum is a main atmospheric characteristic, spanning 0.0071(\rps), corresponding to $\sim9$ atmospheric pressure scale-heights ($H=\frac{kT}{\mu g}=492\pm24$~km, assuming $T=1194^{+58}_{-57}$~K and $g=8.710\pm0.012$) from the optical to the near-infrared (see Fig.\,\ref{fig:cloud_free}).



We followed \cite{lecavelier08a} and performed a fit with a pure scattering model to the transmission spectrum with excluded  sodium and potassium measurements as they can bias the result towards smaller slopes. Assuming an atmospheric opacity source(s) with an effective extinction (scattering$+$absorption) cross section that follows a power law of index $\alpha$, i.e. $\sigma=\sigma_0(\lambda/\lambda_0)^{\alpha}$ the transmission spectrum is then proportional to the product $\alpha T$ given by
\begin{equation}
\alpha T = \frac{\mu g}{k} \frac{{\rm{d}}(\rps)}{{\rm{d}} \ln{\lambda}}.
\label{eq:lecav}
\end{equation}
We find good fit to the 15 \stis\,data points ($\chi^2=5.93$ for $\nu = 13$, $N_{{\rm{free}}}=2$) giving $\alpha T=-3910\pm868$~K.  Adopting the equilibrium temperature\footnote{Although a constant atmospheric temperature is not expected throughout the $\sim9$ pressure scale heights, but is adopted here due to the lack of observations on the planet dayside spectrum.} from \cite{gillon09a}, the slope of the transmission spectrum suggests an effective extinction cross-section of $\sigma=\sigma_0(\lambda/\lambda_0)^{-3.27\pm0.73}$. When fitting with the \sp~\ir~data points we find $\alpha T=-3890\pm576$~K to be a good fit with  ($\chi^2=6.04$ for $\nu=15$, $N_{{\rm{free}}}=2$). In this case  the slope of the transmission spectrum indicates an effective extinction cross-section of $\sigma=\sigma_0(\lambda/\lambda_0)^{-3.26\pm0.48}$. These measurements are in modest disagreement (at the $2.2\text{-}\sigma$ confidence level) with the prediction of Rayleigh scattering from the ground-based result of \cite{Jordan13} who found $\alpha T = - 10670\pm3015$~K. Note that our result significantly improves the error bar of the product $\alpha T$ due to the much broader wavelength coverage. 


\subsubsection{Rayleigh scattering}

When assuming Rayleigh scattering (i.e. adopting $\alpha=-4$), which is the case for a pure gaseous H$_{2}$ atmosphere or scattering by aerosols as in the case of HD 189733b's  atmosphere, the slope implies temperatures of $978\pm217$~K and $973\pm144$~K, when fitting the optical and the complete spectrum, respectively. These temperatures are significantly lower compared to $T=2667\pm750$~K, implied by the analysis of  \cite{Jordan13} assuming Rayleigh scattering. No secondary eclipse observations have been reported in the literature, which prevents an empiric comparison to the planet dayside brightness temperature. However, we can still estimate the expected equilibrium temperature $T_{\rm{eq}}$, which is defined as the surface blackbody temperature for which the incident stellar flux is balanced and is given by:
\begin{equation}
T_{\rm{eq}} =  T_{\ast}  \left( \frac{R_{\ast}}{a} \right)^{1/2} (f (1-A_B))^{1/4},
\end{equation}
where $T_{\ast}$ is the stellar effective temperature, $A_B$ is the bond albedo and $f$ is the redistribution factor. Assuming our best-fit value for \ar, $T_{\ast}=5375\pm65$~K, $f=1/4$ and $A_B=0$ we find $T_{\rm{eq}}=1145\pm23$~K. This value is consistent with our results at the $\sim1\text{-}\sigma$ confidence level.


\begin{figure*}
\centering
\includegraphics[trim = 0 0 0 0, clip, width = 0.95\textwidth]{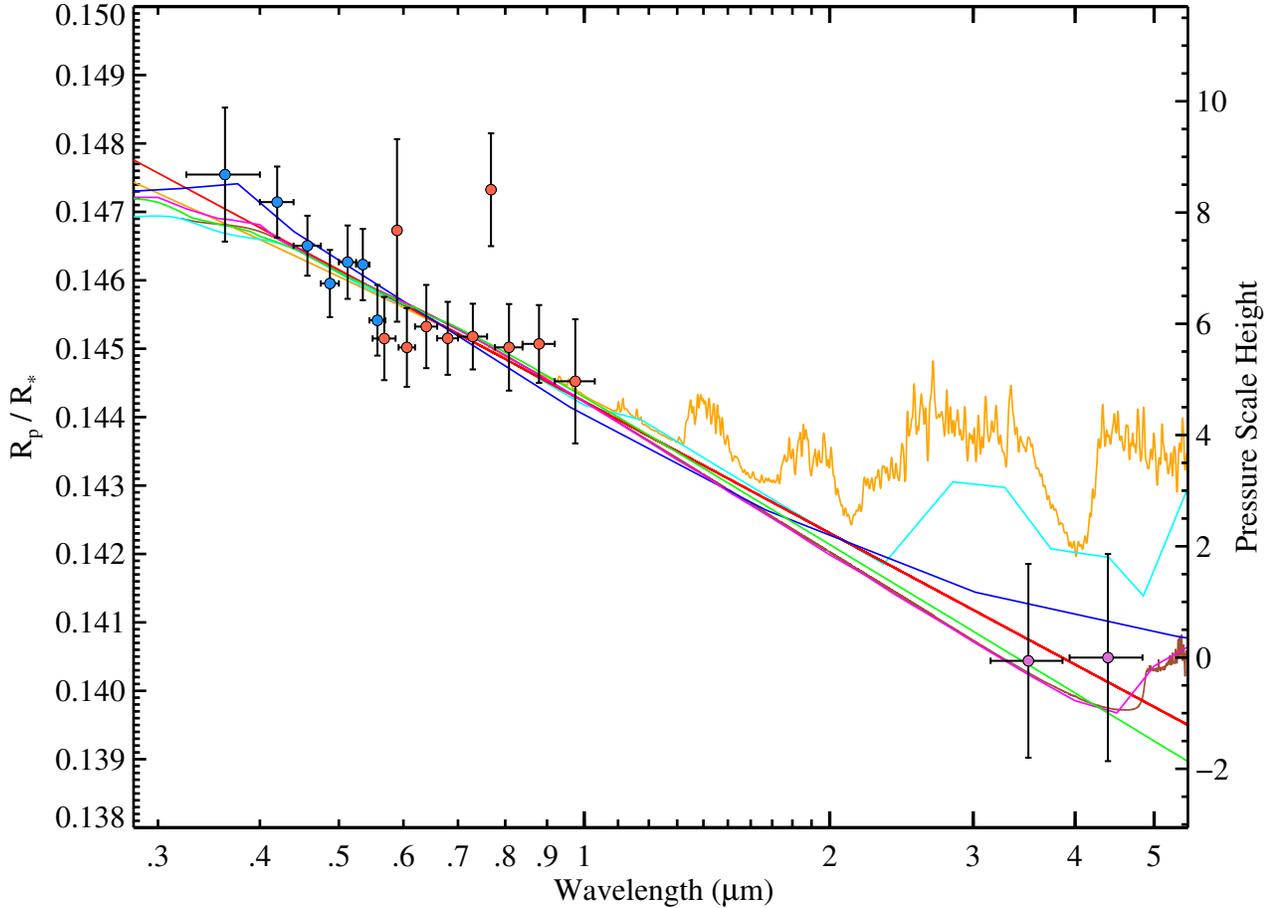}
\caption{ Broad-band transmission spectrum of \wp b compared (without the sodium and potassium measurements)  to seven different aerosol models including:  Rayleigh scattering (red), Mie scattering KCl (green), MgSiO$_3$ (magenta), Fe-poor Mg$_2$SiO$_4$ (brown), a model with enhanced Rayleigh scattering component with a cross-section $10^3$ times that of H$_2$ Fortney.noTiO-EnhancedRayleigh (orange), Na$_2$S (blue) and Titan tholin (cyan).}
\label{fig:transpec_models}
\end{figure*}

\subsubsection{Mie scattering dust}


Good fits to the data are possible assuming a significant opacity from aerosols, i.e. colloids of fine solid particles or liquid droplets in a gas. Aerosol-bearing models assuming Mie theory can fit the observed slope in the spectrum of \wp b. Such materials can be produced from condensate dust species or alternatively from photochemistry \citep{marley13}.  Given the $\sim1000$ K planetary temperatures of \wp b the prime condensate candidate materials include MgSiO$_3$ (enstatite), Mg$_2$SiO$_4$ (forsterite), sulphides and chlorides of the alkali metals Na and K \citep{lodders03, morley12}. Compared to the enstatite and forsterite the alkali sulphides and chlorides are expected to be considerably less abundant in solar-composition atmospheres.

A variety of different atmospheric models were tested assuming aerosols are the main spectral feature throughout the transmission spectrum. The optical properties (i.e. complex indices of refraction as a function of wavelength) were compiled in order of decreasing condensation temperature from 1316 to 90 K for MgSiO$_3$, Fe-poor Mg$_2$SiO$_4$,  Na$_2$S, MnS, KCl, and Titan tholin with all condensates at pressures $10^{-3}$ bar \citep{egan75, zeidler11, khachai09, huffman67, handbook-optical-constants, khare84, ramirez02}. As in \cite{sing13} cross-sections as a function of wavelength were computed with Mie theory which were then used to compute the transmission spectra  using the expression for the planetary altitude derived in  \cite{lecavelier08a}. In all fits we excluded the radius measurements corresponding to the sodium and potassium spectral bins, as these are not predicted by the Mie scattering theoretical models.

First, we assumed a grain size $a$ and planetary temperature equal to the equilibrium temperature and fit for the baseline radius. This approximation is valid given the fact that the cross-section distribution is dominated by the largest grain sizes with $\sigma \propto a^6$ \citep{lecavelier08a}. 

Similar to what was found in \cite{sing13}, when fitting the data for the grain size, temperature and baseline radius simultaneously, a degeneracy between the first two parameters becomes evident. This is not unusual as the slope of the transmission spectrum constrains the quantity $\alpha T$, with the grain size included in the term $\alpha$. In the case of \wp b, we found that Mie scattering model fit with Fe-poor Mg$_2$SiO$_4$ favoured sub-micron grain sizes and higher temperatures ($\sim1400$~K). Similar quality fits with models containing MgSiO$_3$, KCl and Na$_2$S where also found at a temperatures of $1145$ K and grain sizes around $0.1~\mu$m (Fig.~\ref{fig:transpec_models} and Table~\ref{tab:mod_stat2}). The similarity between a pure Rayleigh scattering and Mie scattering models with sub-micron grain sizes from the optical to the near-IR regime is not unusual, as the Mie theory reduces to Rayleigh scattering for particles with sizes much smaller than the wavelength of the light and for compositions with low imaginary components of the index of refraction in the wavelength range considered, as is the case with KCl and MgSiO$_3$.

We also explored the possibility to constrain the condensate and its grain size at lower temperature of 973 K, which we found assuming Rayleigh scattering when fitting the slope of the transmission spectrum. Fitting for the baseline radius and grain size $a$, we found that Fe-poor Mg$_2$SiO$_4$, MgSiO$_3$ and KCl are the most favoured materials with grain size $a=0.1~\mu$m ($\chi^2=6.3$ for $\nu=15$).

We also estimated condensate grain sizes assuming a planetary temperature of 1145 K and fitting for $a$ and the baseline radius. This temperature is similar to the condensation temperature of most of the condensates listed in Table~\ref{tab:mod_stat2}. We find that MgSiO$_3$, KCl, Fe-poor Mg$_2$SiO$_4$ and Na$_2$S are the first four most favoured materials with grain sizes of $\sim0.1~\mu$m.

\begin{table}
\centering
\caption{Model (Mie scattering) selection fit statistics for the complete transmission spectrum ($N=17$).}
\begin{tabular}{@{} l c c c c}
\hline
\hline
Models & $\chi^2$  & $\nu$ & BIC \\
\hline
Rayleigh                       & 6.1   &  15  &  11.7 \\
Mie scattering MgSiO$_3$ or KCl      & 6.7   & 15  &  12.4 \\
Mie scattering Mg$_2$SiO$_4$ (Fe-poor)   & 6.8   &  15  &  12.5 \\
Mie scattering Na$_2$S         & 6.9   & 15  &  12.5 \\
Titan tholin                   & 9.9   & 15  &  15.5 \\
Mie scattering MnS             & 11.4  &  15  &  17.1 \\
Fortney.noTiO-EnhancedRayleigh & 14.0  &  16  &  16.9  \\
Fortney.noTiO-CloudDeck        & 88.2  &  16  &  91.0  \\
\hline
\end{tabular}
\label{tab:mod_stat2}
\end{table}

Finally we examined how our results would change when the optical data is modelled, i.e. excluding the two \ir~measurements. We also excluded the sodium and potassium measurements and fixed the planet temperature to 1145 K and fitted for the grain size and the baseline radius. We find grain size $a\sim0.1~\mu$m for all condensates and the model containing MnS as the most favoured (Table~\ref{tab:mod_stat_optical_only}).

\begin{table}
\centering
\caption{Model fit statistics for the optical spectrum ($N=15$).}
\begin{tabular}{@{} l c c c c}
\hline
\hline
Models & $\chi^2$  & $\nu$ & BIC \\
\hline
Mie scattering MnS                         & 5.1     & 13  &  10.6 \\
Rayleigh                                          & 5.9    & 13  &  11.3 \\
Fortney.noTiO-EnhancedRayleigh  & 6.3    & 14  &  9.0  \\
Mie scattering MgSiO$_3$, Na$_2$S or KCl          & 6.4     & 13  &  11.8 \\
Titan tholin                                      & 6.5    & 13  &  11.9 \\
Mie scattering Mg$_2$SiO$_4$ (Fe-poor)   & 6.6     & 13  &  12.0 \\
Fortney.noTiO-CloudDeck              & 88.2    & 14  &  50.7  \\
\hline
\end{tabular}
\label{tab:mod_stat_optical_only}
\end{table}

\subsubsection{High-altitude haze and lower altitude clouds}
We also compared the transmission spectrum of \wp b with a suite of \cite{fortney10} atmospheric models with either an artificially added Rayleigh scattering component simulating a scattering haze, or an added 'cloud deck' produced by a grey flat line at specific altitude. We performed comparison to these models as they can help better understand our observational results, as clouds and hazes can mask or completely block the expected atomic and molecular features in a transmission spectrum depending on the altitude distribution and particle size. We find that a fit to the data of a 1200 K model without TiO/VO and with a Rayleigh scattering component with a cross-section 1000 times that of H$_2$ gave $\chi^2$ of 14.0 for $\nu=16$. In this model, the optical spectrum is dominated by Rayleigh scattering with H$_2$O features evident in the near-IR (see Fig\,\ref{fig:transpec_models}).

In the Solar System planets, the scale height of haze is often smaller than the scale height of the atmosphere itself, because of sedimentation and the structure of the vertical mixing.  That would result in a smaller Rayleigh slope for the haze signature in the transmission spectrum. It would also add a Ògas-to-grain scale height ratioÓ factor in Eq.1 of \cite{lecavelier08a}, thus severing the relation between the measured slope and the temperature. A typical gas-to-grain scale height ratio in the solar system is 3, which implies a non-negligible correction.

\subsection{The atmosphere of \wp b}

The best-fitting atmospheric models to our data are those containing opacity from Mie or Rayleigh scattering as summarised in Table~\ref{tab:mod_stat1} and \ref{tab:mod_stat2}. All of the cloud-free models have been found to produce unsatisfactory fits indicating low level of molecular absorption in our data. Rayleigh scattering can potentially be produced by molecular hydrogen, a potential scenario explored for HD 209458b \citep{lecavelier08b}. However, molecular hydrogen is disfavoured for \wp b by the lack of pressure-broadened line wings in its transmission spectrum and the low altitudes of any molecular features as apparent in Fig.~\ref{fig:cloud_free}. Rayleigh scattering implies temperatures of $978\pm217$~K or $973\pm144$~K based on a fit to the optical or complete spectrum, respectively. These results are consistent with the estimated planet equilibrium temperature of\,\,$T_{\rm{eq}}=1145\pm23$~K at $\sim1\text{-}\sigma$ confidence level.   

The first five best-fitting Mie scattering aerosol models are practically indistinguishable in quality between each other providing good representation of the observed transmission spectrum (Table~\ref{tab:mod_stat2}). A confident constraint on the composition of the scattering material is hampered by the grain size-temperature degeneracy and the lack of wide spectral coverage. Near-future UV data could potentially resolve these issues, as the Mie scattering models become more distinct in that wavelength region (see Fig.\,\ref{fig:transpec_models}). Despite the degeneracy, we find that the particle sizes of all models are sub-micron, regardless the temperature. On the other hand, near-infrared data such as WFC3 transit spectroscopy or broad-band $JHK$ band transit photometry can bring more evidence for the hazy type of \wp b's atmosphere based on the strength of the water features expected at these regions.

\subsubsection{Comparison to a ground-based result}\label{sec:ground}

Recently, \citet{Jordan13} reported \wp b transmission spectrum with the {\it{Magellan}}~IMACS instrument. With both space- and ground-based results in hand, we are in a position to perform a first rough comparison. We leave a detailed and self-consistent analysis for a future study and simply employ the \hst~\stis data to construct a transmission spectrum using the same spectral bins as in the ground-based study. When fitting the light curves we fixed the system parameters to their best-fitting values. Furthermore, the quality of our data requires us to adopt the four-parameter limb-darkening law rather than the adopted quadratic law for the {\it{Magellan}}~data set. We estimate the uncertainty of each data point in the time series assuming photon noise. 

\begin{figure}
\centering
\includegraphics[trim = 0 0 0 19, clip, width = 0.46\textwidth]{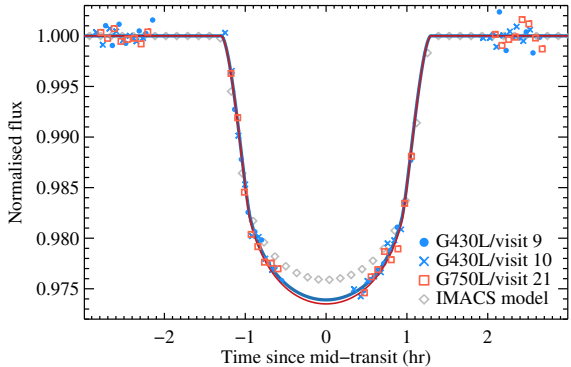}
\caption{\hst~\stis light curves from bin $5468-5647$\,\AA. }
\label{fig:stis_bin_comp}
\end{figure}

Compared to the \hst~transmission spectrum the {\it{Magellan}} spectrum is found to be systematically offset with $\Delta\rps\sim0.007$ lower values. We find no evidence for the observed discrepancy to be due to the different system parameters (i.e. \ar~and $i$) nor the limb darkening laws adopted. However, based on Figure 6 in \citet{Jordan13} where
white light curves are compared after subtracting the best-fit systematics models it can be seen that the final {\it{Magellan}} transit depth varies greatly between the different systematics models used. In particular, one can roughly estimateÊ an average~difference in the transit depth of $\sim0.0017$ between the {\it{white noise}} and {\it{ARMA}} models, which corresponds to a change of $\Delta \rps = 0.0056$. Thus, the differences between the systematics models employed in \citet{Jordan13} could explain more than $\sim80\%$ of the observed offset between the {\it{Magellan}} and \hst~transmission spectra. We therefore find that the potential source of discrepancy between the ground-based study and our results could be primarily linked with the subtracted systematics models of \citet{Jordan13}. Finally, the spectral bin between $5468-5647$~\AA~is an overlapping region for the three \hst~and the {\it{Magellan}} data sets which offers the opportunity for a direct comparison of all data. Fig.\ref{fig:stis_bin_comp} shows that the three \stis\,transit light curves obtained at three individual epochs are all in excellent agreement with each other, illustrating the measured \hst\,transit depth is highly repeatable. 

\begin{figure}
\centering
\includegraphics[trim = 0 0 0 0, clip, width = 0.46\textwidth]{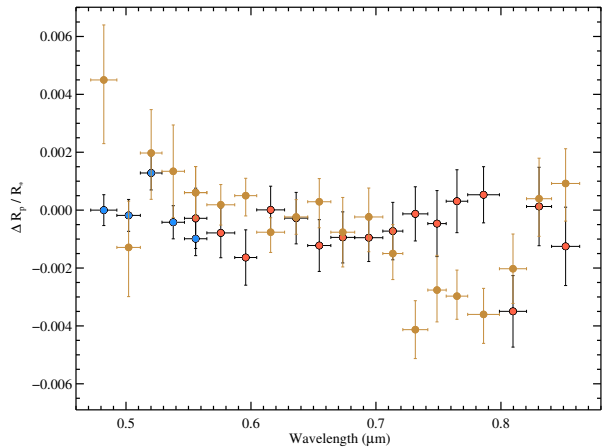}
\caption{Comparison between the \wp b transmission spectrum from \hst~\stis and {\it{Magellan}}~IMACS (blue, red and brown dots, respectively).}
\label{fig:magellan_comp}
\end{figure}

Therefore the comparisons between both results are limited to only the shape of each spectrum, and not the absolute value of the parameter \rps.  When the \hst~and {\it{Magellan}} spectra are normalised to the same average level  (Fig.\,\ref{fig:magellan_comp}), it becomes apparent that there is very good agreement in 15 out of 20 bandpasses, which agree at a $1\text{-}\sigma$ level or less.  Only in 3 of 20 channels one finds a disagreement larger than $2\text{-}\sigma$ (3.1, 2.7, and 3.2$\text{-}\sigma$ in channels  $7215-7415$, $7562-7734$ and $7734-7988$ \AA, respectively).
These regions coincide with several telluric lines (e.g. O$_2$ and H$_2$O), the variation of which on a time interval overlapping with the transit event could potentially introduce variation of the measured depth. The {\it{Magellan}} observation did suffer from changing weather conditions during the second half of the transit event as detailed in \citet{Jordan13}.

\subsubsection{Comparison to HD\,189733b}
Complemented by the transmission spectra of \hdtwo~and \hdone, the first results from the large \hst~program coupled by \sp~IRAC transit radii start building up an observational database of hot-Jupiter atmospheres. The similarities and differences in the spectra and system parameters of the studied objects could potentially indicate distinct categories of gas giant atmospheres. 

\begin{figure}
\includegraphics[trim = 5 1 0 0, clip, width = 0.48\textwidth]{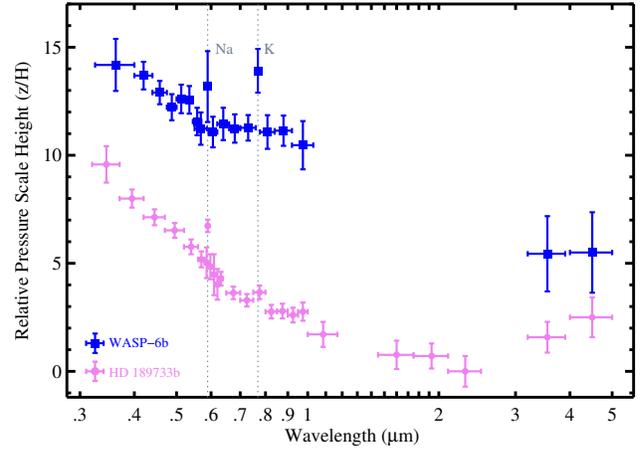}
\caption{Transmission spectrum of \wp b (blue) and HD\,189733b (magenta), taken from \protect\cite{pont13}.}
\label{fig:hd189w6}
\end{figure}

The optical transmission spectrum of \wp b shows some resemblance to that of \hdone~(see~Fig.~\ref{fig:hd189w6}). Both planets also have close zero-albedo equilibrium temperatures near 1200\,K and are hosted by active stars with an overall less activity in \wp~as estimated from HARPS archival data (see Table~\ref{tab:w6_hd189}). Both transmission spectra are characterised by larger optical to near-infrared radii indicating additional scattering in the optical region with featureless slopes and a lack of broad line wings of sodium and potassium which suggests high altitudes and low pressures (e.g. \citealt{desert09, desert11a, pont13} for  \hdone). Instead, absorption from the narrow cores of the Na\,{\sc{i}} and K\,{\sc{i}} doublets are evidenced for \hdone\,in the medium resolution \stis data and the revised ACS data from \citealt{huitson12} and \citealt{pont13}, respectively. In the case of \wp b, we find tentative evidence for narrow core of K\,{\sc{i}}, but no evidence for Na\,{\sc{i}}, which likely is due to the low resolution \stis G750L data\footnote{Table~\ref{tab:specfit_rad} gives the wavelength bins and corresponding radius measurements for the most significant sodium and potassium detections. We note that future detailed direct comparison studies will have to choose a common band system, which potentially can be hampered by the specific resolution and signal-to-noise ratio of each data set.}. These facts are in accord with haze models, as hazes and clouds can reside at high altitudes which can mute or completely block the signatures of atomic and molecular features. Prime condensate candidates responsible for the hazes of \hdone~and \wp b are considered to be silicate condensates such as enstatite \citep{lecavelier08a, pont13}.

Significant differences between both planets and their host stars include a lower effective temperature (i.e. a later spectral type), higher metallicity and a factor of $\sim3$ larger surface gravity of HD189733 and its planet. In addition \hdone's~transmission spectrum exhibits a steeper scattering slope and relatively higher altitude for the hazes on \wp b~(Fig.~\ref{fig:hd189w6}). Together with the planetary equilibrium temperature ($T_{\text{eq}}$), the surface gravity ($g$) is an important parameter in determining the atmospheric scale height and hence the strength of absorption features in exoplanet transmission spectra. A higher surface gravity means that a given pressure corresponds to less mass. Since the clouds usually form at a given pressure level, but the effect of the haze is proportional to the mass, the haze signature in the transmission spectrum is expected to move higher compared to the clouds in a lower-gravity atmosphere (that is part of the reason for instance that Saturn is very hazy while Jupiter is not). That effect goes in the right direction for WASP-6b  vs HD189733b: the first will tend to have a stronger haze signature/deeper cloud deck in transmission. However, with  a factor $\sim3$ in gravity, the difference would be $\sim1$ pressure scale height only, i.e. not enough to account for the full difference between WASP-6b and HD189733b.


Completing the transmission spectra in the near-IR range between 1-3~$\mu$m and improving the \sp~measurement uncertainties will enable a more thorough comparison between the two planets' atmospheres.

\begin{table}
\centering
\caption{System parameters for \wp~and HD\,189733.}
\begin{tabular}{@{} l c c c }
\hline
\hline
Property      &   \wp  &    HD~189733 \\
\hline
$T_{ {\rm{eff}} },\,{\rm{K}}$          &     $5375\pm65$    &  $5050\pm50$  \\
$\log{g} ,\ {\rm{cm\,s^{-2}}}$        &     $4.61\pm0.07$   &  $4.61\pm0.03 $   \\
${\rm{[Fe/H]}}$,\ dex                   &     $-0.20 \pm0.09$  &  $-0.03\pm0.05$   \\

$P, {\rm{day}} $                                           &     $3.361$          &   $2.219$     \\
$a, {\rm{AU}} $                                          &     $0.042$           &   $0.031$       \\
$e        $                                            &   $0$      &     $0$     \\

$g $                                            &     $8.71\pm0.01$      &   $21.5\pm1.2$    \\
$T_{\rm{eq}}$                             &     $1145\pm23$    &   $1191\pm20$        \\
$H,{\rm{km}}$                             &     $492$   &   $199$          \\

\hline
\label{tab:w6_hd189}
\end{tabular}
\end{table}

\section{Conclusion}\label{sec:concl}

With a broad-coverage optical transmission spectrum measured from \hst~and \sp~broad-band transit spectrophotometry, \wp b joins the small but highly valuable family of  hot-Jupiter exoplanets with atmospheric constraints. We observe an overall spectrum characterised by a slope indicative of scattering by aerosols. The data point on the potassium spectral line is $\sim2.7\text{-}\sigma$ higher than the spectrum at that wavelength region. We find no evidence of pressure broadened wings which is in agreement with the haze model as clouds and hazes can reside at high altitudes which can significantly mask or even completely block the signatures of collisionally broadened wings. Given the rather low temperature of \wp b,  aerosol species are expected to originate from high-temperature condensates similar to these in the prototype HD\,189733b. Given their condensation temperatures and the results from our analysis MgSiO$_4$, KCl, MgSiO$_3$ and  Na$_2$S are candidate condensates. 

\wp b is the second planet after \hdone\,which has equilibrium temperatures near $\sim1200$\,K and shows prominent atmospheric scattering in the optical.






  
  


\section*{Acknowledgments}
This work is based on observations with the NASA/ESA {\it{Hubble Space Telescope}}, obtained at the Space Telescope Science Institute (STScI) operated by AURA, Inc. This work is based in part on observations made with the {\it{Spitzer Space Telescope}}, which is operated by the Jet Propulsion Laboratory, California Institute of Technology under a contract with NASA. The research leading to these results has received funding from the European Research Council under the European UnionÕs Seventh Framework Programme (FP7/2007-2013) / ERC grant agreement 336792. NN and DS acknowledge support from STFC consolidated grant ST/J0016/1. PW acknowledge a support from STFC grant. All US-based co-authors acknowledge support from the Space Telescope Science Institute under HST-GO-12473 grants to their respective institutions.

\bibliographystyle{mn2e}
\bibliography{apj-jour,researchv2}

\begin{thebibliography}{84}
\expandafter\ifx\csname natexlab\endcsname\relax\def\natexlab#1{#1}\fi

\bibitem[{{Bean} {et~al}\mbox{.}(2013){Bean}, {D{\'e}sert}, {Seifahrt},
  {Madhusudhan}, {Chilingarian}, {Homeier}, \& {Szentgyorgyi}}]{bean13}
{Bean} J.~L., {D{\'e}sert} J.-M., {Seifahrt} A., {Madhusudhan} N.,
  {Chilingarian} I., {Homeier} D., {Szentgyorgyi} A., 2013, \apj, 771, 108

\bibitem[{{Beerer} {et~al}\mbox{.}(2011){Beerer}, {Knutson}, {Burrows},
  {Fortney}, {Agol}, {Charbonneau}, {Cowan}, {Deming}, {Desert}, {Langton},
  {Laughlin}, {Lewis}, \& {Showman}}]{beerer11}
{Beerer} I.~M. {et~al.}, 2011, \apj, 727, 23

\bibitem[{{Birkby} {et~al}\mbox{.}(2013){Birkby}, {de Kok}, {Brogi}, {de
  Mooij}, {Schwarz}, {Albrecht}, \& {Snellen}}]{birkby13}
{Birkby} J.~L., {de Kok} R.~J., {Brogi} M., {de Mooij} E.~J.~W., {Schwarz} H.,
  {Albrecht} S., {Snellen} I.~A.~G., 2013, \mnras, 436, L35

\bibitem[{{Brogi} {et~al}\mbox{.}(2012){Brogi}, {Snellen}, {de Kok},
  {Albrecht}, {Birkby}, \& {de Mooij}}]{brogi12}
{Brogi} M., {Snellen} I.~A.~G., {de Kok} R.~J., {Albrecht} S., {Birkby} J., {de
  Mooij} E.~J.~W., 2012, \nat, 486, 502

\bibitem[{{Brown} {et~al}\mbox{.}(2001){Brown}, {Charbonneau}, {Gilliland},
  {Noyes}, \& {Burrows}}]{brown01}
{Brown} T.~M., {Charbonneau} D., {Gilliland} R.~L., {Noyes} R.~W., {Burrows}
  A., 2001, \apj, 552, 699

\bibitem[{{Burrows} {et~al}\mbox{.}(2001){Burrows}, {Hubbard}, {Lunine}, \&
  {Liebert}}]{burrows01}
{Burrows} A., {Hubbard} W.~B., {Lunine} J.~I., {Liebert} J., 2001, Reviews of
  Modern Physics, 73, 719

\bibitem[{{Burrows} {et~al}\mbox{.}(2010){Burrows}, {Rauscher}, {Spiegel}, \&
  {Menou}}]{burrows10}
{Burrows} A., {Rauscher} E., {Spiegel} D.~S., {Menou} K., 2010, \apj, 719, 341

\bibitem[{{Burrows} \& {Sharp}(1999)}]{burrows99}
{Burrows} A., {Sharp} C.~M., 1999, \apj, 512, 843

\bibitem[{{Carter} \& {Winn}(2009)}]{carter09}
{Carter} J.~A., {Winn} J.~N., 2009, \apj, 704, 51

\bibitem[{{Charbonneau} {et~al}\mbox{.}(2005){Charbonneau}, {Allen}, {Megeath},
  {Torres}, {Alonso}, {Brown}, {Gilliland}, {Latham}, {Mandushev}, {O'Donovan},
  \& {Sozzetti}}]{charbonneau05}
{Charbonneau} D. {et~al.}, 2005, \apj, 626, 523

\bibitem[{{Charbonneau} {et~al}\mbox{.}(2002){Charbonneau}, {Brown}, {Noyes},
  \& {Gilliland}}]{charbonneau02}
{Charbonneau} D., {Brown} T.~M., {Noyes} R.~W., {Gilliland} R.~L., 2002, \apj,
  568, 377

\bibitem[{{Charbonneau} {et~al}\mbox{.}(2008){Charbonneau}, {Knutson},
  {Barman}, {Allen}, {Mayor}, {Megeath}, {Queloz}, \& {Udry}}]{charbonneau08}
{Charbonneau} D., {Knutson} H.~A., {Barman} T., {Allen} L.~E., {Mayor} M.,
  {Megeath} S.~T., {Queloz} D., {Udry} S., 2008, \apj, 686, 1341

\bibitem[{{Deming} {et~al}\mbox{.}(2013){Deming}, {Wilkins}, {McCullough},
  {Burrows}, {Fortney}, {Agol}, {Dobbs-Dixon}, {Madhusudhan}, {Crouzet},
  {Desert}, {Gilliland}, {Haynes}, {Knutson}, {Line}, {Magic}, {Mandell},
  {Ranjan}, {Charbonneau}, {Clampin}, {Seager}, \& {Showman}}]{deming13}
{Deming} D. {et~al.}, 2013, \apj, 774, 95

\bibitem[{{D{\'e}sert} {et~al}\mbox{.}(2009){D{\'e}sert}, {Lecavelier des
  Etangs}, {H{\'e}brard}, {Sing}, {Ehrenreich}, {Ferlet}, \&
  {Vidal-Madjar}}]{desert09}
{D{\'e}sert} J.-M., {Lecavelier des Etangs} A., {H{\'e}brard} G., {Sing} D.~K.,
  {Ehrenreich} D., {Ferlet} R., {Vidal-Madjar} A., 2009, \apj, 699, 478

\bibitem[{{D{\'e}sert} {et~al}\mbox{.}(2011){D{\'e}sert}, {Sing},
  {Vidal-Madjar}, {H{\'e}brard}, {Ehrenreich}, {Lecavelier Des Etangs},
  {Parmentier}, {Ferlet}, \& {Henry}}]{desert11a}
{D{\'e}sert} J.-M. {et~al.}, 2011, \aap, 526, A12

\bibitem[{{Doyle} {et~al}\mbox{.}(2013){Doyle}, {Smalley}, {Maxted},
  {Anderson}, {Cameron}, {Gillon}, {Hellier}, {Pollacco}, {Queloz}, {Triaud},
  \& {West}}]{doyle13}
{Doyle} A.~P. {et~al.}, 2013, \mnras, 428, 3164

\bibitem[{{Dragomir} {et~al}\mbox{.}(2011){Dragomir}, {Kane}, {Pilyavsky},
  {Mahadevan}, {Ciardi}, {Gazak}, {Gelino}, {Payne}, {Rabus}, {Ramirez}, {von
  Braun}, {Wright}, \& {Wyatt}}]{dragomir11}
{Dragomir} D. {et~al.}, 2011, \aj, 142, 115

\bibitem[{{Eastman} {et~al}\mbox{.}(2010){Eastman}, {Siverd}, \&
  {Gaudi}}]{eastman10}
{Eastman} J., {Siverd} R., {Gaudi} B.~S., 2010, \pasp, 122, 935

\bibitem[{{Egan} \& {Hilgeman}(1975)}]{egan75}
{Egan} W.~G., {Hilgeman} T., 1975, \aj, 80, 587

\bibitem[{{Fazio} {et~al}\mbox{.}(2004){Fazio}, {Hora}, {Allen}, {Ashby},
  {Barmby}, {Deutsch}, {Huang}, {Kleiner}, {Marengo}, {Megeath}, {Melnick},
  {Pahre}, {Patten}, {Polizotti}, {Smith}, {Taylor}, {Wang}, {Willner},
  {Hoffmann}, {Pipher}, {Forrest}, {McMurty}, {McCreight}, {McKelvey},
  {McMurray}, {Koch}, {Moseley}, {Arendt}, {Mentzell}, {Marx}, {Losch},
  {Mayman}, {Eichhorn}, {Krebs}, {Jhabvala}, {Gezari}, {Fixsen}, {Flores},
  {Shakoorzadeh}, {Jungo}, {Hakun}, {Workman}, {Karpati}, {Kichak}, {Whitley},
  {Mann}, {Tollestrup}, {Eisenhardt}, {Stern}, {Gorjian}, {Bhattacharya},
  {Carey}, {Nelson}, {Glaccum}, {Lacy}, {Lowrance}, {Laine}, {Reach},
  {Stauffer}, {Surace}, {Wilson}, {Wright}, {Hoffman}, {Domingo}, \&
  {Cohen}}]{Fazio04}
{Fazio} G.~G. {et~al.}, 2004, \apjs, 154, 10

\bibitem[{{Fortney} {et~al}\mbox{.}(2008){Fortney}, {Lodders}, {Marley}, \&
  {Freedman}}]{fortney08}
{Fortney} J.~J., {Lodders} K., {Marley} M.~S., {Freedman} R.~S., 2008, \apj,
  678, 1419

\bibitem[{{Fortney} {et~al}\mbox{.}(2005){Fortney}, {Marley}, {Lodders},
  {Saumon}, \& {Freedman}}]{fortney05a}
{Fortney} J.~J., {Marley} M.~S., {Lodders} K., {Saumon} D., {Freedman} R.,
  2005, \apjl, 627, L69

\bibitem[{{Fortney} {et~al}\mbox{.}(2010){Fortney}, {Shabram}, {Showman},
  {Lian}, {Freedman}, {Marley}, \& {Lewis}}]{fortney10}
{Fortney} J.~J., {Shabram} M., {Showman} A.~P., {Lian} Y., {Freedman} R.~S.,
  {Marley} M.~S., {Lewis} N.~K., 2010, \apj, 709, 1396

\bibitem[{{Freedman} {et~al}\mbox{.}(2008){Freedman}, {Marley}, \&
  {Lodders}}]{freedman08}
{Freedman} R.~S., {Marley} M.~S., {Lodders} K., 2008, \apjs, 174, 504

\bibitem[{{Gibson} {et~al}\mbox{.}(2013{\natexlab{a}}){Gibson}, {Aigrain},
  {Barstow}, {Evans}, {Fletcher}, \& {Irwin}}]{gibson13a}
{Gibson} N.~P., {Aigrain} S., {Barstow} J.~K., {Evans} T.~M., {Fletcher} L.~N.,
  {Irwin} P.~G.~J., 2013{\natexlab{a}}, \mnras, 428, 3680

\bibitem[{{Gibson} {et~al}\mbox{.}(2013{\natexlab{b}}){Gibson}, {Aigrain},
  {Barstow}, {Evans}, {Fletcher}, \& {Irwin}}]{gibson13b}
{Gibson} N.~P., {Aigrain} S., {Barstow} J.~K., {Evans} T.~M., {Fletcher} L.~N.,
  {Irwin} P.~G.~J., 2013{\natexlab{b}}, \mnras, 436, 2974

\bibitem[{{Gilliland} {et~al}\mbox{.}(1999){Gilliland}, {Goudfrooij}, \&
  {Kimble}}]{gilliland99}
{Gilliland} R.~L., {Goudfrooij} P., {Kimble} R.~A., 1999, \pasp, 111, 1009

\bibitem[{{Gillon} {et~al}\mbox{.}(2009){Gillon}, {Anderson}, {Triaud},
  {Hellier}, {Maxted}, {Pollaco}, {Queloz}, {Smalley}, {West}, {Wilson},
  {Bentley}, {Collier Cameron}, {Enoch}, {Hebb}, {Horne}, {Irwin}, {Joshi},
  {Lister}, {Mayor}, {Pepe}, {Parley}, {Segransan}, {Udry}, \&
  {Wheatley}}]{gillon09a}
{Gillon} M. {et~al.}, 2009, \aap, 501, 785

\bibitem[{{Goudfrooij} \& {Christensen}(1998)}]{goudfrooij98b}
{Goudfrooij} P., {Christensen} J.~A., 1998, {STIS Near-IR Fringing. III. A
  Tutorial on the Use of the IRAF Tasks}. Tech. rep.

\bibitem[{{Grillmair} {et~al}\mbox{.}(2008){Grillmair}, {Burrows},
  {Charbonneau}, {Armus}, {Stauffer}, {Meadows}, {van Cleve}, {von Braun}, \&
  {Levine}}]{grillmair08}
{Grillmair} C.~J. {et~al.}, 2008, \nat, 456, 767

\bibitem[{{Hasan} \& {Bely}(1993)}]{hasan93}
{Hasan} H., {Bely} P.~Y., 1993, in Bulletin of the American Astronomical
  Society, Vol.~25, American Astronomical Society Meeting Abstracts, p. 113.06

\bibitem[{{Hasan} \& {Bely}(1994)}]{hasan94}
{Hasan} H., {Bely} P.~Y., 1994, in The Restoration of HST Images and Spectra -
  II, {Hanisch} R.~J., {White} R.~L., eds., p. 157

\bibitem[{{Henry}(1999)}]{henry99}
{Henry} G.~W., 1999, \pasp, 111, 845

\bibitem[{{Howe} \& {Burrows}(2012)}]{howe12}
{Howe} A.~R., {Burrows} A.~S., 2012, \apj, 756, 176

\bibitem[{Huffman \& Wild(1967)}]{huffman67}
Huffman D.~R., Wild R.~L., 1967, Phys. Rev., 156, 989

\bibitem[{{Huitson} {et~al}\mbox{.}(2013){Huitson}, {Sing}, {Pont}, {Fortney},
  {Burrows}, {Wilson}, {Ballester}, {Nikolov}, {Gibson}, {Deming}, {Aigrain},
  {Evans}, {Henry}, {Lecavelier des Etangs}, {Showman}, {Vidal-Madjar}, \&
  {Zahnle}}]{huitson13}
{Huitson} C.~M. {et~al.}, 2013, \mnras, 434, 3252

\bibitem[{{Huitson} {et~al}\mbox{.}(2012){Huitson}, {Sing}, {Vidal-Madjar},
  {Ballester}, {Lecavelier des Etangs}, {D{\'e}sert}, \& {Pont}}]{huitson12}
{Huitson} C.~M., {Sing} D.~K., {Vidal-Madjar} A., {Ballester} G.~E.,
  {Lecavelier des Etangs} A., {D{\'e}sert} J.-M., {Pont} F., 2012, \mnras, 422,
  2477

\bibitem[{{Husnoo} {et~al}\mbox{.}(2012){Husnoo}, {Pont}, {Mazeh}, {Fabrycky},
  {H{\'e}brard}, {Bouchy}, \& {Shporer}}]{husnoo12}
{Husnoo} N., {Pont} F., {Mazeh} T., {Fabrycky} D., {H{\'e}brard} G., {Bouchy}
  F., {Shporer} A., 2012, \mnras, 422, 3151

\bibitem[{{Jord{\'a}n} {et~al}\mbox{.}(2013){Jord{\'a}n}, {Espinoza}, {Rabus},
  {Eyheramendy}, {Sing}, {D{\'e}sert}, {Bakos}, {Fortney}, {L{\'o}pez-Morales},
  {Maxted}, {Triaud}, \& {Szentgyorgyi}}]{Jordan13}
{Jord{\'a}n} A. {et~al.}, 2013, \apj, 778, 184

\bibitem[{{Katsanis} \& {McGrath}(1998)}]{katsanis98}
{Katsanis} R.~M., {McGrath} M.~A., 1998, {The Calstis IRAF Calibration Tools
  for STIS Data}. Tech. rep.

\bibitem[{Khachai {et~al}\mbox{.}(2009)Khachai, Khenata, Bouhemadou, Haddou,
  Reshak, Amrani, Rached, \& Soudini}]{khachai09}
Khachai H., Khenata R., Bouhemadou A., Haddou A., Reshak A.~H., Amrani B.,
  Rached D., Soudini B., 2009, Journal of Physics: Condensed Matter, 21, 095404

\bibitem[{{Khare} {et~al}\mbox{.}(1984){Khare}, {Sagan}, {Arakawa}, {Suits},
  {Callcott}, \& {Williams}}]{khare84}
{Khare} B.~N., {Sagan} C., {Arakawa} E.~T., {Suits} F., {Callcott} T.~A.,
  {Williams} M.~W., 1984, \icarus, 60, 127

\bibitem[{{Knutson} {et~al}\mbox{.}(2008){Knutson}, {Charbonneau}, {Allen},
  {Burrows}, \& {Megeath}}]{knutson08}
{Knutson} H.~A., {Charbonneau} D., {Allen} L.~E., {Burrows} A., {Megeath}
  S.~T., 2008, \apj, 673, 526

\bibitem[{{Knutson} {et~al}\mbox{.}(2012){Knutson}, {Lewis}, {Fortney},
  {Burrows}, {Showman}, {Cowan}, {Agol}, {Aigrain}, {Charbonneau}, {Deming},
  {D{\'e}sert}, {Henry}, {Langton}, \& {Laughlin}}]{knutson12}
{Knutson} H.~A. {et~al.}, 2012, \apj, 754, 22

\bibitem[{{Lecavelier Des Etangs}
  {et~al}\mbox{.}(2008{\natexlab{a}}){Lecavelier Des Etangs}, {Pont},
  {Vidal-Madjar}, \& {Sing}}]{lecavelier08a}
{Lecavelier Des Etangs} A., {Pont} F., {Vidal-Madjar} A., {Sing} D.,
  2008{\natexlab{a}}, \aap, 481, L83

\bibitem[{{Lecavelier Des Etangs}
  {et~al}\mbox{.}(2008{\natexlab{b}}){Lecavelier Des Etangs}, {Vidal-Madjar},
  {D{\'e}sert}, \& {Sing}}]{lecavelier08b}
{Lecavelier Des Etangs} A., {Vidal-Madjar} A., {D{\'e}sert} J.-M., {Sing} D.,
  2008{\natexlab{b}}, \aap, 485, 865

\bibitem[{{Lewis} {et~al}\mbox{.}(2013){Lewis}, {Knutson}, {Showman}, {Cowan},
  {Laughlin}, {Burrows}, {Deming}, {Crepp}, {Mighell}, {Agol}, {Bakos},
  {Charbonneau}, {D{\'e}sert}, {Fischer}, {Fortney}, {Hartman}, {Hinkley},
  {Howard}, {Johnson}, {Kao}, {Langton}, \& {Marcy}}]{lewis13}
{Lewis} N.~K. {et~al.}, 2013, \apj, 766, 95

\bibitem[{{Lodders}(1999)}]{lodders99}
{Lodders} K., 1999, \apj, 519, 793

\bibitem[{{Lodders}(2002)}]{lodders02a}
{Lodders} K., 2002, \apj, 577, 974

\bibitem[{{Lodders}(2003)}]{lodders03}
{Lodders} K., 2003, \apj, 591, 1220

\bibitem[{{Lodders}(2009)}]{lodders09}
{Lodders} K., 2009, ArXiv:0910.0811

\bibitem[{{Lodders} \& {Fegley}(2002)}]{lodders02b}
{Lodders} K., {Fegley} B., 2002, \icarus, 155, 393

\bibitem[{{Lodders} \& {Fegley}(2006)}]{lodders06}
{Lodders} K., {Fegley}, Jr. B., 2006, {Chemistry of Low Mass Substellar
  Objects}, p.~1

\bibitem[{{Mandel} \& {Agol}(2002)}]{mandel02}
{Mandel} K., {Agol} E., 2002, \apjl, 580, L171

\bibitem[{{Markwardt}(2009)}]{markwardt09}
{Markwardt} C.~B., 2009, in Astronomical Society of the Pacific Conference
  Series, Vol. 411, Astronomical Data Analysis Software and Systems XVIII,
  {Bohlender} D.~A., {Durand} D., {Dowler} P., eds., p. 251

\bibitem[{{Marley} {et~al}\mbox{.}(2013){Marley}, {Ackerman}, {Cuzzi}, \&
  {Kitzmann}}]{marley13}
{Marley} M.~S., {Ackerman} A.~S., {Cuzzi} J.~N., {Kitzmann} D., 2013, {Clouds
  and Hazes in Exoplanet Atmospheres}, {Mackwell} S.~J., {Simon-Miller} A.~A.,
  {Harder} J.~W., {Bullock} M.~A., eds., pp. 367--391

\bibitem[{{Mighell}(2005)}]{mighell05}
{Mighell} K.~J., 2005, \mnras, 361, 861

\bibitem[{{Morley} {et~al}\mbox{.}(2012){Morley}, {Fortney}, {Marley},
  {Visscher}, {Saumon}, \& {Leggett}}]{morley12}
{Morley} C.~V., {Fortney} J.~J., {Marley} M.~S., {Visscher} C., {Saumon} D.,
  {Leggett} S.~K., 2012, \apj, 756, 172

\bibitem[{{Nikolov} {et~al}\mbox{.}(2013){Nikolov}, {Chen}, {Fortney},
  {Mancini}, {Southworth}, {van Boekel}, \& {Henning}}]{nikolov13}
{Nikolov} N., {Chen} G., {Fortney} J.~J., {Mancini} L., {Southworth} J., {van
  Boekel} R., {Henning} T., 2013, \aap, 553, A26

\bibitem[{{Nikolov} {et~al}\mbox{.}(2014){Nikolov}, {Sing}, {Pont}, {Burrows},
  {Fortney}, {Ballester}, {Evans}, {Huitson}, {Wakeford}, {Wilson}, {Aigrain},
  {Deming}, {Gibson}, {Henry}, {Knutson}, {Lecavelier des Etangs}, {Showman},
  {Vidal-Madjar}, \& {Zahnle}}]{nikolov14}
{Nikolov} N. {et~al.}, 2014, \mnras, 437, 46

\bibitem[{{O'Rourke} {et~al}\mbox{.}(2014){O'Rourke}, {Knutson}, {Zhao},
  {Fortney}, {Burrows}, {Agol}, {Deming}, {D{\'e}sert}, {Howard}, {Lewis},
  {Showman}, \& {Todorov}}]{orourke14}
{O'Rourke} J.~G. {et~al.}, 2014, \apj, 781, 109

\bibitem[{Palik(1998)}]{handbook-optical-constants}
Palik E.~D., 1998, Handbook of Optical Constants of Solids. Elsevier

\bibitem[{{Pont} {et~al}\mbox{.}(2008){Pont}, {Knutson}, {Gilliland}, {Moutou},
  \& {Charbonneau}}]{pont08}
{Pont} F., {Knutson} H., {Gilliland} R.~L., {Moutou} C., {Charbonneau} D.,
  2008, \mnras, 385, 109

\bibitem[{{Pont} {et~al}\mbox{.}(2013){Pont}, {Sing}, {Gibson}, {Aigrain},
  {Henry}, \& {Husnoo}}]{pont13}
{Pont} F., {Sing} D.~K., {Gibson} N.~P., {Aigrain} S., {Henry} G., {Husnoo} N.,
  2013, \mnras, 432, 2917

\bibitem[{{Ramirez} {et~al}\mbox{.}(2002){Ramirez}, {Coll}, {da Silva},
  {Navarro-Gonz{\'a}lez}, {Lafait}, \& {Raulin}}]{ramirez02}
{Ramirez} S.~I., {Coll} P., {da Silva} A., {Navarro-Gonz{\'a}lez} R., {Lafait}
  J., {Raulin} F., 2002, \icarus, 156, 515

\bibitem[{{Reach} {et~al}\mbox{.}(2005){Reach}, {Megeath}, {Cohen}, {Hora},
  {Carey}, {Surace}, {Willner}, {Barmby}, {Wilson}, {Glaccum}, {Lowrance},
  {Marengo}, \& {Fazio}}]{reach05}
{Reach} W.~T. {et~al.}, 2005, \pasp, 117, 978

\bibitem[{{Redfield} {et~al}\mbox{.}(2008){Redfield}, {Endl}, {Cochran}, \&
  {Koesterke}}]{redfield08}
{Redfield} S., {Endl} M., {Cochran} W.~D., {Koesterke} L., 2008, \apjl, 673,
  L87

\bibitem[{Schwarz(1978)}]{schwarz78}
Schwarz G., 1978, Annals of Statistics, 6, 461

\bibitem[{{Seager} \& {Sasselov}(2000)}]{seager00}
{Seager} S., {Sasselov} D.~D., 2000, \apj, 537, 916

\bibitem[{{Sharp} \& {Burrows}(2007)}]{sharp07}
{Sharp} C.~M., {Burrows} A., 2007, \apjs, 168, 140

\bibitem[{{Sing}(2010)}]{sing10}
{Sing} D.~K., 2010, \aap, 510, A21

\bibitem[{{Sing} {et~al}\mbox{.}(2012){Sing}, {Huitson}, {Lopez-Morales},
  {Pont}, {D{\'e}sert}, {Ehrenreich}, {Wilson}, {Ballester}, {Fortney},
  {Lecavelier des Etangs}, \& {Vidal-Madjar}}]{sing12}
{Sing} D.~K. {et~al.}, 2012, \mnras, 426, 1663

\bibitem[{{Sing} {et~al}\mbox{.}(2013){Sing}, {Lecavelier des Etangs},
  {Fortney}, {Burrows}, {Pont}, {Wakeford}, {Ballester}, {Nikolov}, {Henry},
  {Aigrain}, {Deming}, {Evans}, {Gibson}, {Huitson}, {Knutson}, {Showman},
  {Vidal-Madjar}, {Wilson}, {Williamson}, \& {Zahnle}}]{sing13}
{Sing} D.~K. {et~al.}, 2013, \mnras, 436, 2956

\bibitem[{{Sing} {et~al}\mbox{.}(2011){Sing}, {Pont}, {Aigrain}, {Charbonneau},
  {D{\'e}sert}, {Gibson}, {Gilliland}, {Hayek}, {Henry}, {Knutson}, {Lecavelier
  Des Etangs}, {Mazeh}, \& {Shporer}}]{sing11b}
{Sing} D.~K. {et~al.}, 2011, \mnras, 416, 1443

\bibitem[{{Sing} {et~al}\mbox{.}(2008){Sing}, {Vidal-Madjar}, {D{\'e}sert},
  {Lecavelier des Etangs}, \& {Ballester}}]{sing08a}
{Sing} D.~K., {Vidal-Madjar} A., {D{\'e}sert} J.-M., {Lecavelier des Etangs}
  A., {Ballester} G., 2008, \apj, 686, 658

\bibitem[{{Snellen} {et~al}\mbox{.}(2008){Snellen}, {Albrecht}, {de Mooij}, \&
  {Le Poole}}]{snellen08}
{Snellen} I.~A.~G., {Albrecht} S., {de Mooij} E.~J.~W., {Le Poole} R.~S., 2008,
  \aap, 487, 357

\bibitem[{{Snellen} {et~al}\mbox{.}(2010){Snellen}, {de Kok}, {de Mooij}, \&
  {Albrecht}}]{snellen10a}
{Snellen} I.~A.~G., {de Kok} R.~J., {de Mooij} E.~J.~W., {Albrecht} S., 2010,
  \nat, 465, 1049

\bibitem[{{Stevenson} {et~al}\mbox{.}(2014){Stevenson}, {Bean}, {Seifahrt},
  {D{\'e}sert}, {Madhusudhan}, {Bergmann}, {Kreidberg}, \&
  {Homeier}}]{stevenson14}
{Stevenson} K.~B., {Bean} J.~L., {Seifahrt} A., {D{\'e}sert} J.-M.,
  {Madhusudhan} N., {Bergmann} M., {Kreidberg} L., {Homeier} D., 2014, \aj,
  147, 161

\bibitem[{{Suchkov} \& {Hershey}(1998)}]{suchkov98}
{Suchkov} A., {Hershey} J., 1998, {NICMOS Focus and HST Breathing}. Tech. rep.

\bibitem[{{Todorov} {et~al}\mbox{.}(2013){Todorov}, {Deming}, {Knutson},
  {Burrows}, {Fortney}, {Lewis}, {Cowan}, {Agol}, {Desert}, {Sada},
  {Charbonneau}, {Laughlin}, {Langton}, \& {Showman}}]{todorov13}
{Todorov} K.~O. {et~al.}, 2013, \apj, 770, 102

\bibitem[{{Visscher} {et~al}\mbox{.}(2006){Visscher}, {Lodders}, \&
  {Fegley}}]{visscher06}
{Visscher} C., {Lodders} K., {Fegley}, Jr. B., 2006, \apj, 648, 1181

\bibitem[{{Wakeford} {et~al}\mbox{.}(2013){Wakeford}, {Sing}, {Deming},
  {Gibson}, {Fortney}, {Burrows}, {Ballester}, {Nikolov}, {Aigrain}, {Henry},
  {Knutson}, {Lecavelier des Etangs}, {Pont}, {Showman}, {Vidal-Madjar}, \&
  {Zahnle}}]{wakeford13}
{Wakeford} H.~R. {et~al.}, 2013, \mnras, 435, 3481

\bibitem[{{Werner} {et~al}\mbox{.}(2004){Werner}, {Roellig}, {Low}, {Rieke},
  {Rieke}, {Hoffmann}, {Young}, {Houck}, {Brandl}, {Fazio}, {Hora}, {Gehrz},
  {Helou}, {Soifer}, {Stauffer}, {Keene}, {Eisenhardt}, {Gallagher}, {Gautier},
  {Irace}, {Lawrence}, {Simmons}, {Van Cleve}, {Jura}, {Wright}, \&
  {Cruikshank}}]{Werner04}
{Werner} M.~W. {et~al.}, 2004, \apjs, 154, 1

\bibitem[{{Zeidler} {et~al}\mbox{.}(2011){Zeidler}, {Posch}, {Mutschke},
  {Richter}, \& {Wehrhan}}]{zeidler11}
{Zeidler} S., {Posch} T., {Mutschke} H., {Richter} H., {Wehrhan} O., 2011,
  \aap, 526, A68

\end{thebibliography}
\end{document}